\documentclass[a4paper,11pt]{article}

\usepackage{jheppub} 

\usepackage[T1]{fontenc} 

\title{Quartic Horndeski, planar black holes, holographic aspects and universal bounds}

\author[a,1]{Jos\'e Pablo Figueroa}
\author[b,2]{Konstantinos Pallikaris}

\affiliation[a]{Departamento de F\'{\i}sica, Universidad de Concepci\'{o}n, Casilla, 160-C, Concepci\'{o}n, Chile.}
\affiliation[b]{Laboratory of Theoretical Physics, Institute of Physics, University of Tartu, W. Ostwaldi 1, 50411 Tartu, Estonia.}

\emailAdd{josepfigueroa@udec.cl}
\emailAdd{konstantinos.pallikaris@ut.ee}

\abstract{In this work, we consider a specific shift-invariant quartic Horndeski model, deriving new planar black hole solutions with axionic hair. We explore these solutions in terms of their horizon structure and their thermodynamic properties. We use the gauge/gravity dictionary to derive the DC transport coefficients of the holographic dual with the aim of investigating how the new deformation affects the universality of some renown bound proposals. Although most of them are found to hold true, we nevertheless find a highly interesting parametric violation of the \emph{heat conductivity-to-temperature} lower bound which acquires a dependence on both the scale and the coupling. Finally, using a perturbative approach, a more brutal violation of the \emph{viscocity-to-entropy} ratio is demonstrated.}

\usepackage{hyperref}
\usepackage{caption}
\usepackage{graphicx}
\usepackage{dcolumn}
\usepackage{bm}
\usepackage{cancel}
\usepackage{physics}
\usepackage{xcolor}
\usepackage{dsfont}
\usepackage{stmaryrd}
\usepackage{upgreek}
\usepackage{subfigure}

\begin{document} 
\maketitle
\flushbottom

\section{Introduction}

Scalar-tensor theories of gravity have been well studied in the past, their pros and cons, as alternative theories of gravitation, elucidated in detail. They enrich the dynamical field content of general relativity by the inclusion of scalar fields in the latter, which constitute additional degrees of freedom. One of the most renowned scalar-tensor family is Horndeski gravity~\cite{Horndeski:1974wa}, the most general four-dimensional scalar-tensor theory with equations of motion containing up to second-order derivatives of the dynamical fields. The field content of Horndeski gravity consists of the spacetime metric $g_{\mu\nu}$ and a scalar field $\phi$. We focus on the subclass where the latter enjoys a global shift symmetry under which $\phi\to \phi+c$, with $c$ being some constant. In this scenario, the Horndeski action involves four arbitrary functions of the canonical kinetic term $X:=-(\partial\phi)^2/2$, denoted by $G_i$, $i=2,...,5$. It reads
\begin{equation}
S=\int d^{\,4}x\sqrt{-g}\mathcal{L}=:\int d^{\,4}x\sqrt{-g}\sum_{i=2}^{5}\mathcal{L}_i,\label{eq:HornAction}
\end{equation}
where
\begin{eqnarray}
\mathcal{L}_2&=&G_2,\quad \mathcal{L}_3=-G_3\Box\phi,\quad \mathcal{L}_4=G_4R+G_{4,X}\bqty{(\Box\phi)^2-(\phi_{\mu\nu})^2},\nonumber\\
\mathcal{L}_5&=&G_5G_{\mu\nu}\phi^{\mu\nu}-\frac{G_{4,X}}{6}\bqty{(\Box\phi)^3-3(\phi_{\mu\nu})^2\Box\phi+2(\phi_{\mu\nu})^{3}}.
\end{eqnarray}
Here, the following shorthand notation is used: $\phi_{\mu\nu. . .}:=\nabla_\mu\nabla_\nu. . .\phi$ and 
\begin{equation}
(\phi_{\mu\nu})^{p}:=\phi^{\lambda_1}{}_{\lambda_2}\phi^{\lambda_2}{}_{\lambda_3}. . .\phi^{\lambda_p}{}_{\lambda_1}.
\end{equation}
Additionally, $R$ is the Ricci scalar, and $G_{\mu\nu}=R_{\mu\nu}-\frac{1}{2}g_{\mu\nu}R$ is the Einstein tensor. In the context of covariant Galileon theory, these nonminimal couplings play the role of counterterms which cancel out with the higher derivative terms arising from the variation of the action~\cite{Dvali:2000hr,Nicolis:2008in,Deffayet:2009wt}. 

As said above, the shift symmetry enjoyed by the scalar field is an artefact of the restriction $G_i(\phi,X)\to G_i(X)$ which assures that {the action  \eqref{eq:HornAction} is not algebraic in $\phi$}. Due to this symmetry, it becomes possible to write the equation of motion for $\phi$, i.e., the Klein-Gordon equation, in terms of the Noether current associated with global shift symmetry, 
\begin{equation}
\partial_\mu(\sqrt{-g}J^{\mu})=0,\quad J^{\mu}=-\fdv{\mathcal{L}}{\phi_\mu},\quad \fdv{S}{\phi}=-\partial_\mu\fdv{S}{\phi_\mu}.
\end{equation}
Usually, when introducing additional degrees of freedom, one tries to see whether the latter allow for solutions with nontrivial profiles. No-hair theorems such as~\cite{Ruffini:1971bza,Bekenstein:1995un} state assumptions which, if met, forbid {any deviation} from the solution spectrum of general relativity. More relevant to our case, are the no-hair arguments of~\cite{Hui:2012qt} applying to static and spherically symmetric spacetimes in the framework of Horndeski gravity. By using the $\mathrm{SO}(3)$ symmetry of the ansatz and the time-reversal invariance of the action, the authors of~\cite{Hui:2012qt} show that the only nonvanishing component of the aforementioned current is the radial one. Then, assuming (i) asymptotic flatness and (ii) regularity of diffeo-invariant quantities at the horizon, like $J_\mu J^{\mu}$, together with (iii) vanishing boundary conditions at infinity --- that is, taking $\phi'$ to vanish there ---, it is finally proven that $J^{r}(r_0)=0$ where
\begin{equation}
J^{r}=g^{rr}\phi' H(\phi',g,g',g'').
\end{equation}
This leads to $J^{r}=0$ everywhere via the conservation law, followed by the conclusion $\phi'=0$ at all radii, provided {that $H$ asymptotically tends} to a nonzero constant when $\phi'\to 0$. Following the refinement in~\cite{Babichev:2017guv}, the assumptions have to be supplemented with {two extra arguments: (iv) the functions $G_i$ should be chosen in such a way} that their derivatives with respect to $X$ do not introduce negative powers of $X$ as the latter approaches the origin, and (v) the canonical kinetic term $X$ must be present in the action. Then, the theorem guarantees that static, spherically symmetric black holes with nontrivial scalar field profiles cannot exist. 

One controversial beauty of no-hair theorems revolves around possible ways of circumventing their prohibitive results. Indeed, it has been shown that relaxing some of the hypotheses of~\cite{Hui:2012qt} allows for nontrivial scalar hair. Giving up on (i), several (A)dS or Lifshitz black holes were reported~\cite{Rinaldi:2012vy,Anabalon:2013oea,Cisterna:2014nua,Minamitsuji:2013ura,Kobayashi:2014eva, Bravo-Gaete:2013dca}. In these cases, although $H$ asymptotically tends to 0, the scalar field profile is nevertheless nontrivial. Another circumventing route goes through allowing the scalar field to linearly depend on time~\cite{Babichev:2013cya}, providing several stealth solutions with asymptotic behaviors depending on whether (v) is violated or not. These solutions have also provided a fruitful ground for the construction of neutron stars which avoid conflicts with solar system tests~\cite{Cisterna:2015yla,Cisterna:2016vdx}. Furthermore, failing (iv), static hairy solutions with asymptotic flatness can be constructed~\cite{Babichev:2017guv}. The lesson to be learned here is that by relaxing the hypotheses, one either strengthens no-hair theorems or avoids their confining results.

{Here, we will consider the action proposed in~\cite{Babichev:2017guv}, namely $G_3=G_5=0$ and $G_2\sim X$, $G_4\sim \sqrt{-X}$. This has a firm motivation when a radial profile is assumed for the scalar field; the Horndeski functions are chosen in such a way so as to meet condition (v), but fail condition (iv), respectively, in an attempt to obtain hairy solutions. However, here we will add a twist to it which is interesting from a holographic point of view, as explained in the next paragraph. Instead of a radial scalar field profile, we will assume a dependence of the scalar field on the spatial coordinates $\{x^{i}\}$, $i=1,...,D-2$, where $D$ is the spacetime dimension. The latter comprise a choice of local chart on a Euclidean submanifold of dimension $D-2$ corresponding to the spatial piece of the boundary. To be able to derive a black hole solution, we will have to introduce a collection of $(D-2)$-many such scalar fields, $\{\psi^I\}$, which we homogeneously distribute along the spatial directions, adopting the ansatz $\psi^I\equiv \psi^I(x^i)$. To this end, $(D-2)$-many copies of the aforementioned Horndeski model will be considered, one for each scalar field. Here, $I=1,...,D-2$ stands for an internal index used for labeling these fields.} We will focus on {scalar field} solutions breaking translation symmetry in the planar directions, i.e., massless  St\"uckelberg fields with a linear bulk profile  $\psi^{I}=p\delta^{I}_ix^{i}$. They can be equally understood as magnetically charged 0-forms with their charge being proportional to the slope $p$ of the profile; in this sense, they should classify as primary hair~\cite{Caldarelli:2016nni}. We will work our way towards a new family of electrically charged hairy planar black holes characterized by a nonzero axion background.\footnote{See~\cite{Cai:1996eg,Cai:1997ii} for earlier works on plane solutions.} In general, the idea of looking for black hole solutions with various types of scalar fields or $k$-form fields homogeneously distributed along the planar directions, has been a fruitful practice as evidenced by some articles, e.g., see~\cite{Bardoux:2012aw, Bardoux:2010sq, Caldarelli:2013gqa}. {Particularly, in the midst of other highly interesting proposals in~\cite{Bardoux:2012aw}, the inclusion of scalar fields --- one for each coordinate of the horizon submanifold --- with a linear spatial profile in the bulk action, was utilised to  shape new axionic black hole solutions with AdS asymptotics and a planar horizon geometry.}

{As previously argued, on top of obtaining a novel family of exact solutions, considering such a bulk profile for the scalars promotes another agenda which is interesting in its own right, when viewed  through the prism of holography: boundary momentum gets dissipated resolving the delta function multiplying the Drude weight in the conductivity formula; dissipation happens exactly because these massless scalars --- and hence the leading terms in their boundary expansion which source marginal operators deforming the CFT --- have a spatial dependence.} Simple boundary $\mathrm{U}(1)$- and diffeomorphism-symmetry arguments suffice to derive the diffeomorphism Ward identity for the (non)conservation of the boundary stress tensor,
\begin{equation}
\nabla_{\alpha}\ev{T^{\alpha\beta}}=\ev{\mathcal{O}_I}\nabla^{\beta}\psi_{(0)}^{I}+F^{\beta}_{(0)\alpha}\ev{\mathcal{J}^\alpha},
\end{equation}
where one eventually sees that $\ev{T^{tx^i}}$ fails to be conserved, exactly due to the spatial dependence of $\psi^{I}_{(0)}$. {Here, Greek indices from the start of the Greek alphabet are used to label components $\{t,x^i\}$.} The presence of this relaxation mechanism will allow us to pursue the second objective of this work which is to apply holographic techniques in order to compute the DC transport coefficients~\cite{Andrade:2013gsa,Jiang:2017imk,Donos:2014cya} of the holographic dual in the broader gauge/gravity duality context of the renown AdS/CFT correspondence~\cite{Maldacena:1997re,Witten:1998qj}. Many studies in GR and alternative theories of gravitation have been carried out, and their results, from a holographic point of view, have been elucidated~\cite{Caldarelli:2016nni,Bardoux:2012tr,Kuang:2018ymh,Kuang:2017rpx,Kuang:2017cgt,Feng:2017jub,Cisterna:2017jmv,Cisterna:2018hzf,Cisterna:2019uek}. Furthermore, gauge/gravity duality has been also providing towards the study of fascinating phenomena in strongly correlated systems, indicative examples being the linear $T$-resistivity and the universal Homes's law~\cite{Hartnoll:2014lpa,Hartnoll:2016apf,Erdmenger:2015qqa,Kim:2015dna}.\footnote{See~\cite{Cremonini:2018kla,Kiritsis:2016cpm} for holographic strange metals.} Insight from holography has also been given into various bounds and their possible universality, examples being the {electric and} thermoelectric conductivity bounds~\cite{Grozdanov:2015djs,Grozdanov:2015qia}, the Kovtun-Son-Starinets (KSS) bound~\cite{Kovtun:2004de,Policastro:2001yc} of the viscosity-to-entropy ratio, the universal bounds for the charge/energy diffusion constants~\cite{Hartnoll:2014lpa,Blake:2016wvh,Blake:2016sud} in the regime where diffusive physics dominate e.t.c. Especially interesting was the refinement~\cite{Blake:2016wvh} of the original Hartnoll proposal~\cite{Hartnoll:2014lpa} --- the latter supported by experimental data on dirty metals as well --- by identifying the characteristic velocity of the system with the so-called butterfly velocity $v_B$~\cite{Roberts:2014isa,Shenker:2013pqa,Geng:2020kxh}, which measures how fast quantum information scrambles and proves to be a good candidate at strong coupling. As the refinement is an outcome of holographic methods, it always makes sense to probe it in various holographic models. In general, certain deformations of the bulk action can affect the universality of some of these bounds; higher derivative corrections~\cite{Baggioli:2016pia} can drive the charge diffusitivity bound all the way to zero, the inclusion of the Gauss-Bonnet term lowers the \emph{viscosity-to-entropy} ratio to a smaller $\mathcal{O}(1)$ number~\cite{Brigante:2007nu}, {various violations of the electric conductivity bound were reported in~\cite{Baggioli:2016oqk,Baggioli:2016oju,Gouteraux:2016wxj}, either by coupling the scalar fields, or their kinetic terms, directly to the Maxwell term, or via a nonlinear generalization of standard Maxwell electrodynamics, e.t.c. Motivated by these interesting facets,  and by the holographic treatment of scalar-tensor theories, e.g., see refs.~\cite{Baggioli:2017ojd,Garcia-Garcia:2016hsd,Jiang:2017imk}, we are prompted to investigate these quantities in our case, where a violation of the conjectured bounds is a plausible scenario due to a Horndeski-type deformation, the strength of the latter controlled by the coupling constant of the theory.} 

This paper is organized as follows: In section~\ref{sec:2} we formulate the action principle extracting its variational field equations. We derive electrically charged planar black holes  with axion hair in arbitrary spacetime dimensions $D>3$, followed by a discussion about the horizon structure of the four-dimensional solution. We close the section by presenting a straightforward extension for dyons, while we also display the three-dimensional case which cannot be derived by simply taking the limit. Section~\ref{sec:thermo} is dedicated to the study of the thermodynamic properties of the four-dimensional plane solution with AdS asymptotics, where we derive the entropy via both the Wald formalism and the Euclidean path integral approach. Both methods agree on the result, and the first law is shown to hold true {provided that the parameter controlling the strength of boundary momentum dissipation is held fixed}. The 1/4-area law for the entropy is modified, although it can be said to hold in the broader sense, in units of an effective gravitational coupling~\cite{Brustein:2007jj}. We move on to section~\ref{sec:Holo} where we probe holographic features of the bulk theory starting with DC transport coefficients. After we derive the thermoelectric response of the dual system, we compute the Lorentz ratios, proceeding with a discussion about the lower bound of heat conductivity, $\kappa$. {Due to the new coupling, we find an explicit violation of the bound, whereas we also show that, at fixed temperature and chemical potential, the new lower bound depends on the amount of dissipation and the strength of the gravity bulk deformation}. Next, in order to probe the proposed diffusitivity bounds, we first show that in the strong dissipation regime, regardless of the coupling strength, the mixed diffusion constants, $D_{\pm}$, decouple as the mixing term becomes negligible, and one can use the simpler formulas for the charge and energy diffusion, $D_c$ and $D_e$, respectively. {After determining the butterfly velocity, $v_B$, we demonstrate that the nonminimal coupling does not have a qualitative impact on the incoherence-dominated physics; the $D_{c(e)}/v_B^2$ ratios are bounded by the standard numbers from below.} Neither does the proposed deformation play any role in the low temperature expansion. In general, it does not have any leading-order contribution in these cases. {Finally, we end the section with a very shallow investigation of the \emph{viscosity-to-entropy} ratio via perturbative methods where an explicit violation of the simple KSS bound is manifest as expected, with its ``brutality'' driven by the new coupling. In section~\ref{sec:conc} we summarize our results and conclude. }

\section{The model: action, field equations and hairy solutions\label{sec:2}}
Let us start by introducing the model we will work on. Let $X^{I}=-(\partial \psi^{I})^2/2$ be the canonical kinetic term of the $I$-th scalar field.\footnote{Summation for repeated internal indices is not assumed, unless otherwise stated.} Then, the shorthand notation $G^{I}_\bullet:=G_\bullet(X^{I})$ will be convenient. Also, when writting $\psi^{I}_{\mu\nu. . .}$, we mean $\nabla_\mu\nabla_\nu. . .\psi^{I}$. {Having said that, we restrict ourselves to the quartic sector of \eqref{eq:HornAction} with $G_3=0=G_5$ and 
\begin{equation}
G^I_2=\hat{\eta} X^I-\frac{2\Lambda}{(D-2)},
\end{equation}
for $\hat{\eta}\in \mathds R$ and in $16\pi \mathrm{G_N}=1$ units. We then choose 
\begin{equation}
G^I_4=\alpha\sqrt{-X^I}+\frac{1}{D-2}.
\end{equation}
All of these choices boil down to the following Lagrangians: 
\begin{equation}
\mathcal{L}_2^{I}=G_2^I,\quad \mathcal{L}_4^{{I}}=\frac{ R}{(D-2)}+\alpha\sqrt{-X^{{I}}}R-\frac{\alpha}{2\sqrt{-X^{{I}}}}\Psi^{I},
\end{equation} 
where ${\Psi^{I}}:=(\Box \psi^{I})^2-(\psi^{I}_{\mu\nu})^2$.} We consider the action functional
\begin{equation}
S=\int d^Dx\sqrt{-g}\sum_{I=1}^{D-2}\sum_{n=2}^{5}\mathcal{L}_n^{{I}}-\frac{1}{4}\int d^{D}x\sqrt{-g}\mathcal F^2,\label{eq:toymodel}
\end{equation}
with $\mathcal F^2:=F^{\mu\nu}F_{\mu\nu}$ and $F_{\mu\nu}=2\partial_{[\mu}A_{\nu]}$, $A_\mu$ being the $\mathrm{U}(1)$ field. {The necessary surface terms accompanying~\eqref{eq:toymodel} are omitted, yet implied.} In a more clear form, the principle \eqref{eq:toymodel} can be rewritten as 
\begin{equation}
S=S_{\mathrm{GR}}+\sum_{{I}=1}^{D-2}S_{\psi^{{I}}}+S_{\mathrm{EM}},
\end{equation}
where
\begin{eqnarray}
S_{\mathrm{GR}}&=&\int d^{D}x\sqrt{-g}\pqty{R-2\Lambda},\\
S_{\psi^{{I}}}&=&\int d^{D}x\sqrt{-g}\pqty{\hat{\eta} X^{{I}}+\alpha\sqrt{-X^{{I}}}R-\frac{\alpha}{2\sqrt{-X^{{I}}}}\Psi^{I}},\\
S_{\mathrm{EM}}&=&-\frac{1}{4}\int d^{D}x\sqrt{-g}\mathcal F^2.
\end{eqnarray}
Observe here that $S_{\psi^{{I}}}$ contains the nonminimal coupling as well.

{As anticipated in the introduction, the action has a structural form similar to the principle in~\cite{Babichev:2017guv}; nevertheless, here we consider $(D-2)$-many copies of the Horndeski piece, the latter constructed out of scalar fields with a spatial profile, in order to derive a new exact plane solution as advertised in the beginning. The piece of the action containing second derivatives of the metric reads
\begin{equation}
    \int d^Dx\sqrt{-g}\pqty{1+\alpha \sum_{I}\sqrt{-X^I}}R.
\end{equation}
For now, the dimensionless coupling constant $\alpha$ is left largely unrestricted; it only has to be real. Moreover, $\hat \eta$ is a parameter of length dimension -2, determining the sign of the kinetic terms. We will in general demand $\hat\eta >0$ so that when $\alpha$ is switched off, we recover the Einstein-Maxwell theory supplemented by kinetic terms for the scalar fields which come with the correct sign. Correct in the sense  that the reduced model is devoid of ghosts. Notice also that there exists another interesting feature of the Horndeski Lagrangian. Neglecting the Einstein-Hilbert term, the action enjoys global scale symmetry under the transformations $g_{\mu \nu} \rightarrow \lambda^{2} g_{\mu \nu} $ and $\psi^I \rightarrow \psi^I/\lambda$, exactly as shown in \cite{Padilla:2013jza}. }

Stationary variations with respect to $g^{\mu\nu}$, $A_\mu$ and $\psi^{{I}}$ yield
\begin{equation}
\delta S=\int d^{D}x\sqrt{-g}\bqty{\mathcal{G}_{\mu\nu}\var g^{\mu\nu}+\nabla_\nu F^{\nu\mu}\var A_\mu+\sum_{{I}=1}^{D-2}\nabla^{\mu}J^{{I}}_\mu\delta\psi^{{I}}},\label{eq:varS}
\end{equation}
where for convenience we have defined 
\begin{equation}
    \mathcal{G}_{\mu\nu}:={\mathcal{G}^{\mathrm{GR}}_{\mu\nu}+\sum_{{I}=1}^{D-2} \mathcal G_{\mu\nu}^{\psi^{{I}}}+\mathcal G^{\mathrm{EM}}_{\mu\nu}}.\label{eq:Gtot}
\end{equation}
Here, 
\begin{eqnarray}
\mathcal{G}^{\mathrm{GR}}_{\mu\nu}&=&G_{\mu\nu}+g_{\mu\nu}\Lambda,\label{eq:GGR}\\
\mathcal{G}^{\mathrm{EM}}_{\mu\nu}&=&-\frac{1}{2}\pqty{F_{\mu\rho}F_{\nu}{}^{\rho}-\frac{1}{4}g_{\mu\nu}\mathcal F^2}\label{eq:GEM},
\end{eqnarray}
while
\begin{eqnarray}
\mathcal{G}^{\psi^{{I}}}_{\mu\nu}\!&=&\!-\frac{\hat{\eta}}{2}\pqty{\psi^{{I}}_\mu\psi^{{I}}_\nu+g_{\mu\nu}X^{{I}}}+\alpha\sqrt{-X^{{I}}}G_{\mu\nu}-\nonumber\\
&&\!+\frac{G_{4,X^{{I}}}^{{I}}}{2}\bqty{2\pqty{R_{\mu}{}^{\rho}{}_\nu{}^{\sigma}\!-\!g_{\mu\nu}R^{\rho\sigma}}\psi^{{I}}_{\rho}\psi^{{I}}_{\sigma}\!+\!4R^{\rho}{}_{(\mu}\psi^{{I}}_{\nu)}\psi^{{I}}_{\rho}\!-\!R\psi^{{I}}_{\mu} \psi^{{I}}_\nu\!}+\nonumber\\
&&\!+\frac{G_{4,X^{{I}}}^{{I}}}{2}\Bqty{\!2\pqty{\psi^{{I}}_{\mu}{}^{\rho}\psi_{\rho\nu}^{{I}}\!-\!\psi^{{I}}_{\mu\nu}\Box\psi^{{I}}}\!+\!g_{\mu\nu}{\Psi^{I}} }\!\!+\!\nonumber\\
&&\!+\frac{G_{4,X^{{I}}X^{{I}}}^{{I}}}{2}\Bqty{2\pqty{g_{\mu\nu}\psi^{{I}}_{\rho}{}^{\lambda}\psi_{\lambda\sigma}^{{I}}\psi^{I\rho}\psi^{I\sigma}-\psi_{\rho}^{{I}}\psi^{{I}}_{\mu}{}^\rho\psi_{\sigma}^{{I}}\psi^{{I}}_{\nu}{}^\sigma}-{\Psi^I}\psi^{{I}}_\mu\psi^{{I}}_\nu}-\nonumber\\
&&\!-G_{4,X^{{I}}X^{{I}}}^{{I}}\bqty{2\psi^{I}_{\rho(\mu}\psi^{{I}}_{\nu)}-\psi^{I}_{\rho}\psi^{{I}}_{\mu\nu}+\pqty{g_{\mu\nu}\psi^{{I}}_\rho-2g_{\rho(\mu}\psi^{{I}}_{\nu)}  }\Box\psi^{{I}}}\psi^{I\sigma}\psi^{{I}}_{\sigma}{}^\rho.\label{eq:Gpsi}
\end{eqnarray}
whereas
\begin{equation}
J^{{I}}_\mu=\hat{\eta}\psi^{{I}}_\mu-2G_{4,X^{{I}}}^{{I}}G_{\mu\nu}\psi^{{I}\nu}+G_{4,X^{{I}}X^{{I}}}^{{I}}\pqty{{\Psi^I}\psi^{{I}}_\mu -2\psi^{{I}}_{\nu}\psi^{{I}}_{\mu}{}^\nu\Box\psi^{{I}}+2\psi^{{I}}_{\rho}\psi^{I}_{\nu}{}^\rho\psi^{{I}}_{\mu}{}^{\nu}}.
\end{equation}
In what follows, we will focus on a particular class of solutions to the Klein-Gordon equations, namely $\psi^{{I}}=p\delta^I_ix^{i}$; this breaks translation invariance in the planar directions but retains the little $\mathrm{SO}(D-2)$ symmetry. Such a solution will also significantly simplify the calculations.

\subsection{Electrically charged planar black holes with a nontrivial axion profile}
Let us consider a static spherically symmetric metric,
\begin{equation}
ds^2=-F(r)dt^2+\frac{dr^2}{F(r)}+r^2\delta_{ij}dx^{i}dx^{j},\label{eq:ssymansatz}
\end{equation}
together with the bulk profile $\psi^{{I}}\equiv \psi^I(x)$, and let us start solving equations. The easiest equation to start with is the Maxwell one. We consider a purely electric field strength tensor $F_{\mu\nu}=-\mathcal{A}'(r)\delta^{tr}_{\mu\nu}$ for a Maxwell potential 1-form $A=\mathcal{A}(r)dt$. Then, we need to satisfy
\begin{equation}
-\partial_r\pqty{r^{D-2}\mathcal{A}'(r)}=0.
\end{equation}
This is solved by an electric field
\begin{equation}
\mathcal{A}'(r)=\frac{Q_e}{r^{D-2}}\label{eq:Efield}
\end{equation}
Let us proceed with the Klein-Gordon {equation for the $I$-th scalar field}. A specific solution to this is $\psi^I\equiv \psi^i =px^i$. To see that this is a solution, notice that first of all, $X^{{i}}=-p^2/(2r^2)$. Second, the only nonvanishing component of $J^{I\mu}$ is the $i$-th one, 
\begin{equation}
J^{Ix^{{i}}}=\pqty{\frac{\hat{\eta} p}{r^2}+\frac{\alpha\Bqty{(D-3)\bqty{(D-4)F+2rF'}+r^{2}F''}}{\sqrt{2}r^3}}\delta^{Ii},\label{eq:axioncurr}
\end{equation}
where {in order to write down this expression, we used the explicit expression of the $x^ix^i$ component of the Einstein tensor (not to be confused with $\mathcal{G}_{\mu\nu}$) in the ansatz \eqref{eq:ssymansatz}, which reads}

\begin{equation}
G_{x^ix^{{i}}}=\frac{1}{2}\Bqty{(D-3)\bqty{(D-4)F+2rF'}+r^2F''}.
\end{equation}

It is evident that the Klein-Gordon {equation for the $I$-th scalar field,  
\begin{equation}
\partial_\mu(\sqrt{-g}J^{I\mu})=\partial_{x^i}(r^2J^{I x^i})=0,
\end{equation}
is identically solved, because eq. \eqref{eq:axioncurr} --- the only nontrivial component of the $I$-th current --- is a function of the radius only. Having satisfied the gauge and scalar field equations, we move on to the metric field equations. Here, we just solve one component, say $\mathcal{G}_{tt}$, the solution being
\begin{equation}
F(r)=\frac{r}{{2 r\!+\!\sqrt{2}(D-3)\alpha \abs{p}}}\pqty{\frac{Q_e^2}{(D-2)(D-3)r^{2(D-3)}}\!-\!\frac{M}{r^{D-3}}\!-\!\frac{\hat{\eta} p^2}{D-3}\!-\!\frac{4\Lambda r^2}{(D-1)(D-2)}}.\label{eq:arbDsol}
\end{equation}
The above metric function satisfies all the remaining components of the metric field equations, always given the electric field \eqref{eq:Efield} and the axion profile $\psi^i=px^i$.} Moreover, notice that as we continuously approach $\alpha=0$, the limit where the nonminimal coupling vanishes and the theory reduces to the Einstein-Maxwell-(linear) Axion (EMA) model, the solution expands as 
\begin{equation}
F(r)=-\frac{M}{2 r^{D-3}}-\frac{2\Lambda r^2}{(D-1)(D-2)}+\frac{Q_e^2}{2(D-2)(D-3) r^{2(D-3)}}-\frac{\hat{\eta} p^2}{2(D-3)}+\mathcal{O}(\alpha),
\end{equation}
i.e. the electrically charged axionic solution in~\cite{Bardoux:2012aw} is recovered as expected. The latter can also be found in~\cite{Andrade:2013gsa}. {Finally, we impose the constraint $\alpha >0$ so that \eqref{eq:arbDsol} and the curvature invariants are analytic everywhere, except at the origin which corresponds to an inescapable pole. }

\subsection{The four-dimensional solution\label{sec:4Dsol}}
The four-dimensional solution is given by the $D\to 4$ limit of \eqref{eq:arbDsol}. It reads
{\begin{equation}
F(r)=\frac{r}{2 r+\sqrt{2}\alpha \abs{p}}\pqty{\frac{Q_e^2}{2r^2}-\frac{M}{r}-\hat{\eta} p^2 -\frac{2\Lambda r^2}{3}},\label{eq:f4}
\end{equation}}
and admits the asymptotic expansion
{
\begin{equation}
F(r)=-\frac{\Lambda r^2}{3}+\frac{\alpha \abs{p} \Lambda r}{3\sqrt{2}}-\frac{p^2(3\hat{\eta}+\alpha^2\Lambda) }{6}-\frac{M_{\mathrm{eff}}}{ r}+\frac{Q^2_{\mathrm{eff}}}{4r^2}+\mathcal{O}(1/r^3),
\end{equation}
where the effective mass and the effective charge read
\begin{equation}
M_{\mathrm{eff}}:=\frac{M}{2}-\frac{\alpha \abs{p}^3(3\hat{\eta}+\alpha^2\Lambda)}{6\sqrt{2}},\qquad Q^2_{\mathrm{eff}}:=Q_e^2+2\sqrt{2}\alpha \abs{p} M_{\mathrm{eff}}.
\end{equation}
First, this is the most general solution;} Had we started with $g_{tt}=-U(r)$ in the ansatz, we would see that the system of field equations would force $U$ to be a multiple of $F$ times an integration constant which can always be fixed such that $U=F$. {Some further comments are in order here. In the chosen region of the parameter space, defined by $\alpha >0$ and $\hat{\eta}>0$, the theory can only accommodate planar black hole solutions with AdS asymptotics. Otherwise, the singularity is naked. However, we have fixed the sign of $\hat \eta$ by demanding that our theory is continuously reduced to the EMA model when taking the limit $\alpha\to 0$. We shall however investigate the sign options a bit more. }

{To do this, let us consider a diagonal vierbein satisfying the orthonormality relation $g_{\mu\nu}=\eta_{ab}e^a_\mu e^b_\nu$ where $\eta=\mathrm{diag}(-1,1,1,1)$, $g_{\mu\nu}$ is given by~\eqref{eq:ssymansatz}, and $F$ is as in~\eqref{eq:f4}. We can write the metric field equations as $\mathcal{G}_{\mu\nu}^{\mathrm{GR}}= T_{\mu\nu}$ where we have defined the effective stress tensor of the bulk theory,  $ T_{\mu\nu}:=\mathcal{G}_{\mu\nu}^{\mathrm{GR}}-\mathcal{G}_{\mu\nu}$. The involved quantities can be found in Eq. \eqref{eq:Gtot}--\eqref{eq:Gpsi}. The effective energy density, $\rho\equiv \rho(r)$, is then given by $\rho=-T^0{}_0=-T^t{}_te^0{}_t e_0{}^t=-T^t{}_t$. We find that 
\begin{equation}
    \rho=\frac{Q_e^2+2\hat\eta p^2r^2+2\sqrt{2}\alpha \abs{p}F' r^2}{4r^4},
\end{equation}
where $F$ and its derivatives are to be treated on the shell. This expands as
\begin{equation}
\rho=-\frac{\sqrt{2}\alpha \abs{p}\Lambda}{3r}+\mathcal{O}(1/r^2),
\end{equation}
when $r$ approaches infinity, whereas
\begin{equation}
    \rho=\frac{Q_e^2+2  r_0^2(\hat{\eta}p^2-\sqrt{2}\alpha \abs{p} \Lambda r_0)}{2r_0^3(2r_0+\sqrt{2}\alpha \abs{p})}+\mathcal{O}(r-r_0),
\end{equation}
when expanding about the event horizon. First of all, since we want a positive $\rho$ in the physical domain $r\in [r_0,+\infty)$, it turns out that for $\Lambda<0$, even when $\hat\eta$ is negative, but bounded from below as 
\begin{equation}
    0 > \hat\eta > \frac{\sqrt{2}\alpha \Lambda r_0}{\abs{p}},
\end{equation}
we still have a positive energy density. For flat and de Sitter asymptotics, the choice $\hat \eta<0$ unavoidably leads to regions of negative density in the respective, physically sensible, radial domains; ergo, we will not consider these cases, although we mention for completion that for $\hat\eta<0$, it is possible to construct an asymptotically flat black hole with two horizons. In what follows, we will stick to the limit argument, as expressed in the end of the previous paragraph. This means $\hat{\eta}>0$.
}

{From now on, let us also fix $\Lambda=-3$ once and for all setting the radius of AdS to unity}. The horizons of \eqref{eq:f4} are located at the real positive roots of the depressed quartic equation
\begin{equation}
r^4-\frac{\hat{\eta} p^2}{2}r^2-\frac{ M}{2}r+\frac{Q_e^2}{4}=0.\label{eq:horeq}
\end{equation}
The multiplicity, as well as the reality of the roots, depends on the sign of the discriminant. It is best if we study the extrema of the auxiliary function
\begin{equation}
W(r):=(2r+\sqrt{2}\alpha \abs{p})F+M,
\end{equation}
instead. These are located at the positive real solutions of 
\begin{equation}
r^4-\frac{\hat{\eta} p^2}{6}r^2-\frac{Q_e^2}{12}=0.
\end{equation}
We find that there exists only one positive real solution which corresponds to the global minimum
\begin{equation}
M_*:=W(r_*)=\frac{12Q_e^2-\hat{\eta} p^2\mathcal{C}}{3\sqrt{3\mathcal{C}}},\qquad \mathcal{C}:=\hat{\eta} p^2+\sqrt{\hat{\eta}^2p^4+{12Q_e^2}},\label{eq:extremhor}
\end{equation}
located at $r_*=\sqrt{\mathcal C}/(2\sqrt{3})$ which is independent of the new parameter $\alpha$. For $M<M_*$ the singularity is naked, whereas when $M>M_*$ there exist two horizons, the outer one being the event horizon $r_0$ located at the largest root of \eqref{eq:horeq}, and the inner one being a Cauchy horizon. We remark that their location can be analytically determined since eq. \eqref{eq:horeq} is analytically solvable, but the explicit expressions are too lengthy to write down. When $M=M_*$, an extremal black hole forms, with its horizon located at $r_0=r_*$, the latter also being written as
\begin{equation}
r_\ast=\frac{\sqrt{2\hat{\eta} p^2+\mu^2}}{2\sqrt{3}},\label{eq:exthor}
\end{equation}
in terms of what will later be identified with the chemical potential of the holographic dual, given the expression $\mathcal{A}=\mu(1-r_0/r)$. The extremal black hole goes to a unit-radius $\text{AdS}_4$ at infinity, while near the horizon, an $\text{AdS}_2\times \mathds R^{2}$ product structure appears with 
\begin{equation}
L_{\text{AdS}_2}^2=\frac{2\hat{\eta} p^2+\mu^2+\sqrt{6}\alpha \abs{p} \sqrt{2\hat{\eta} p^2 +\mu^2}}{6(\hat{\eta} p^2+\mu^2)},
\end{equation}
which of course agrees with the findings in~\cite{Andrade:2013gsa}, that is when $\alpha=0$. 

{Note that $M_*$ can in theory be negative, so we can also have black holes with a negative mass parameter.} {As pointed out in~\cite{Bardoux:2012aw}, the magnitude of the negative mass black hole is strongly tied to the magnitude of the axionic charge via a proportionality relation. Indeed, from \eqref{eq:extremhor} we see that since $\mathcal{C}>0$ always, negative masses are allowed for positive $\hat{\eta}$ only, driven by the relative strength of the axionic charge, the negativity condition being
\begin{equation}
    r_\ast >\frac{\abs{Q_e}}{\sqrt{\hat{\eta}}\abs{p}}.
\end{equation}
Things are more lucid in the absence of an electric charge, where the minimum mass bound $M_*$ is always negative. This case does exactly reveal the proportionality relation which reads $M_*\propto -\abs{p}^3$. In particular, when this is the case, we only have two horizons in the negative mass region
\begin{equation}
0>M>-\frac{\sqrt{2}}{3\sqrt{3}}\abs{p}^3\eta^{3/2}=:M_*,\label{eq:masextnoQ}
\end{equation}
where an extremal black hole forms for $M=M_*$ at $r_*=\abs{p}\sqrt{\hat\eta/6}$. When $M<M_*$, the singularity is naked, whereas when $M>0$ there exists only one horizon veiling a true curvature singularity at $r=0$.
}

{Interestingly enough, there also exists a black hole solution with $M=Q_e=0$. In this setup, the metric function reads
\begin{equation}
    {F}(r)=\frac{r}{2 r+\sqrt{2}\alpha \abs{p}}\pqty{2r^2-\hat\eta p^2 }.\label{eq:nomasnocharge}
\end{equation}
It has a real positive root at $r_0=\abs{p}\sqrt{\hat{\eta}/2}$ which stands for the location of the event horizon, the latter sourced by the axionic charge. To see that this hides a singularity at the origin, we display the Ricci scalar, which near $r=0$ expands as 
\begin{equation}
R=\frac{3\sqrt{2}\hat\eta \abs{p}}{\alpha r}+\mathcal{O}(1),
\end{equation}
exhibiting a simple pole at the origin. The existence of the pole is not due to the new coupling; even in the absence of the nonminimal coupling, the Ricci scalar would still be singular at the origin since then, $R=\hat\eta p^2/r^2 -12$. Note that eq. \eqref{eq:nomasnocharge} is the $Q_e\to 0$ limit of the vanishing mass version of \eqref{eq:f4}, the latter representing a charged hairy planar black hole in AdS with an inner Cauchy horizon and an outer event horizon, both of them coalescing into a single horizon at $r_*=\abs{p}\sqrt{\hat\eta}/2$ when the black hole becomes extremal. These results are well studied in~\cite{Bardoux:2012aw} for minimally coupled axion fields. Although our case differs due to the presence of the Horndeski coupling, we see that extremality conditions are not affected in general. Moreover, in the absence of mass and charges, the theory admits a planar AdS vacuum, although, if one treats $p$ as a fixed nonzero parameter, then it is reasonable to consider the extremal black hole with $Q_e=0$ as the vacuum of the theory, since its euclidean version describes a regular spacetime in the valid domain.}  We find it is also worth mentioning that the dyonic extension of \eqref{eq:f4}, for a Maxwell potential 1-form $A=\mathcal{A}(r)dt+Q_m x^{[1}dx^{2]}$, is straightforward. The solution representing a dyonic black hole simply reads
\begin{equation}
F(r)=\frac{r}{2 r+\alpha \abs{p}\sqrt{2}}\pqty{\frac{Q_e^2+Q_m^2}{2r^2}-\frac{M}{r}-\hat{\eta} p^2  +2r^2},\label{eq:f4dyon}
\end{equation}
where the expected interchange duality $Q_e\leftrightarrow Q_m$ is apparent. We would like to close this subsection by displaying the three-dimensional solution as well. The latter corresponds to a logarithmic branch and it cannot follow from the $D\to 3$ limit of \eqref{eq:arbDsol}. In this separate case,
\begin{equation}
F(r)=-M+ r^2-\frac{Q_e^2+\hat{\eta} p^2}{2\kappa}\ln r,
\end{equation}
and it is evident that this simply is a charged BTZ solution with the axion flux playing the role of the magnetic charge. 

\section{Black hole thermodynamics\label{sec:thermo}}

In this section, we will focus on the thermodynamic properties of the black hole solutions derived in Sec. \ref{sec:4Dsol}. Indeed, 
even if \eqref{eq:f4} has the standard AdS asymptotic behavior, it nevertheless remains interesting to investigate it in terms of black hole thermodynamics. This study is further motivated by the presence of an
unusual coupling between the scalar curvature and the square root of the kinetic term. Such a coupling is
expected to modify the 1/4-area law of the entropy as we will see below. On the other hand, as it was pointed out in~\cite{Feng:2015oea}, the presence 
of a nonminimal coupling generates some obscure facets when analyzing the thermodynamic properties of static Horndeski black holes. Indeed, in the last reference, an
asymptotically AdS static black hole solution of a particular $G_2$- and $G_4$-Horndeski Lagrangian~\cite{Anabalon:2013oea} was scrutinized from a thermodynamical point of view. It was observed that the Wald formalism~\cite{Feng:2015oea}, the regularized Euclidean method~\cite{Anabalon:2013oea} and  the quasilocal approach~\cite{Peng:2015yjx}, all applied to this specific solution, give
rise to distinct expressions of the thermodynamic quantities. This
is somehow intriguing since these different approaches are usually
consistent with each other.\footnote{See the recent work~\cite{Hajian:2020dcq} for  fresh insight, as well as a resolution proposal.} These discrepancies
can essentially be attributed to the nonminimal coupling between the geometry
and the derivatives of the scalar field, but also to the fact that the
static scalar field and its radial derivative diverge at the
horizon. {Here, in our case, such a behaviour is not present because there is no radial profile for the scalar field at all; the latter is radially constant in this sense. Nevertheless, it remains interesting to investigate if the presence of the nonminimal coupling, alone, can source discrepancies like the ones mentioned above. In the remainder of this work, we will consider $p>0$ without loss of generality.}

As a first step, we will compute the so-called Wald entropy, ${\cal S}_{\mathrm W}$, defined by
{
\begin{eqnarray}
{\cal S}_{\mathrm W}=-2\pi \int \fdv{{\cal L}}{ R_{\mu\nu\rho\sigma}}\epsilon_{\mu\nu}\epsilon_{\rho\sigma}\bar{\epsilon}=-8\pi r_0^2\int d^2x\fdv{\mathcal{L}}{R_{trtr}}=\hat\sigma\left(4  \pi r_0^2+4\sqrt{2}\pi\alpha {p} r_0\right),
\label{eq:WaldEntropy}
\end{eqnarray}
for the solution \eqref{eq:f4}.} Here, the integral is taken over a slice of the horizon and ${\cal L}$ is the full Lagrangian. Also, $\epsilon_{\mu\nu}$ denotes the unit bivector, normal to the horizon surface, while $\bar{\epsilon}$ stands for the area of the slice. Finally, $\hat\sigma$ denotes the volume of the planar base sub-manifold. As previously anticipated, the nonminimal coupling between the
scalar curvature and the square root of the kinetic term does indeed modify the standard 1/4-area law of the entropy. If $\mathcal S_o$ is the standard entropy $4\hat\sigma\pi r_0^2$ in $16\pi G_{\mathrm{N}}=1$ units, then Wald's Noether charge entropy formula simply spits out
\begin{equation}
\mathcal S_{\mathrm W}=\mathcal S_o\pqty{1+\frac{\sqrt{2}\alpha {p}}{r_0}}.\label{eq:entroalt}
\end{equation}
In the sense of~\cite{Brustein:2007jj}, the 1/4-area law still holds in $16\pi G_{\mathrm{eff}}=1$ units where the effective running gravitational coupling takes the particular value
\begin{equation}
G_{\mathrm{eff}}:=\frac{1}{16\pi}\pqty{1+\frac{\sqrt{2}\alpha {p}}{r}}^{-1}_{r=r_0},\label{eq:kappaef}
\end{equation}
an expression that could have been equally guessed from \eqref{eq:toymodel} as well. We will also confirm \eqref{eq:WaldEntropy} by means of the
Euclidean approach for which the thermal partition function is identified with the Euclidean path integral at the saddle point around the classical solution. 

To do so, we consider the following Euclidean ansatz:
\begin{eqnarray}
ds^2=N(r)^2\,F(r)d\tau^2+\frac{dr^2}{F(r)}+r^2d\Sigma_{2,\gamma=0},\qquad \psi^{{i}}=\psi^{{i}}(x^{{i}}),\qquad x^{i}=\{x,y\},
\label{EuclAnsatz}
\end{eqnarray}
together with an electric ansatz $A_{\mu}dx^{\mu}=\mathcal A(r)d\tau$. Here, $\tau$ is the Euclidean periodic time with period $0\leq \tau<T^{-1}=:\beta$ where $T$ stands for the Hawking temperature. The range of the radial
coordinate $r$ is given by $r\geq r_0$. In the mini superspace of the symmetry ansatz \eqref{EuclAnsatz}, the
Euclidean action, $I_{E}$, {reads
\begin{eqnarray}
I_E\!=\!\beta\int dr d^2x N\Bqty{-6 r^2\!+\!2 F\!+\!\frac{\hat{\eta}}{2}\bqty{(\psi^{1}_{x})^2\!+\!(\psi^{2}_{y})^2} \!+\!F'\bqty{2 r \!+\!\frac{\alpha}{\sqrt{2}}  { \pqty{\abs{\psi^{1}_{x}}\!+\!\abs{\psi^{2}_{y}}} }  }}\!&-&\nonumber\\
-{\beta}\int dr d^2x\pqty{\frac{N\Pi^2}{2r^2}+\mathcal{A}\Pi'}+B_E.&&
\label{EuclAction}
\end{eqnarray}
Here, ${\Pi}:=N r^2 F^{rt}$ denotes the radial momentum, canonically conjugate to the gauge field.} Moreover, $B_E$ is an appropriate boundary term codifying all the thermodynamic properties, while also ensuring that
the solution corresponds to an extremum of the action, i.e., $\var I_E=0$. Note that the volume element of the Euclidean action is not only radial, as it usually is. This is due to the fact that the scalar fields are assumed to depend on the planar coordinates $x$ and $y$. A simple exercise shows that the Euler-Lagrange equations obtained from variation of the symmetry-reduced action \eqref{EuclAction} with respect to the dynamical fields $F$, $N$,  $\mathcal A$ and $\psi^{{i}}$ lead to the {euclidean version} of the solution \eqref{eq:f4} with $N=\text{const.}$, which we can set to unity without loss of generality. 

We now consider the formalism of the grand canonical ensemble, varying the Euclidean action while keeping fixed the temperature, the electric potential and the parameter $p$ controlling the strength of momentum dissipation in the dual theory. Under these considerations, the extremality condition $\var I_E=0$ fixes the variation of the boundary term {to read
\begin{eqnarray}
\var B_E=-\hat\sigma\beta  \left\lbrack N\left(2 r+\frac{\alpha\sqrt{2}}{2}\left({ \abs{\psi^{1}_{x}}\!+\!\abs{\psi^{2}_{y}} }\right)\right)\var F-\mathcal A\var{\Pi}\right\rbrack_{r=r_0}^{r=\infty},\label{boundvar}
\end{eqnarray}
where $\hat \sigma$ is the volume of the two-dimensional compact planar submanifold.} At infinity, eq. \eqref{boundvar} evaluates to $\delta B_E(\infty)=\hat\sigma\beta\pqty{\var M+\mu\var Q_e}$, while at the horizon, the lack of conical singularity --- ensured by requiring that
$\var F(r_0)=-4\pi T\var r_0$ --- {yields
\begin{equation}
\var B_E(r_0)=4\pi \hat \sigma\beta T\left(2 r_0+\sqrt{2}\alpha p\right)\var r_0.
\end{equation}
Then,
\begin{eqnarray}
B_E&=&\hat\sigma\beta\bqty{ \pqty{M-M_*}-T\left(4 \pi r_0^2+4\sqrt{2}\pi\alpha p r_0\right)+\mu Q_e},
\label{bterm}
\end{eqnarray}
where we have set the arbitrary constant, which can in theory appear in the boundary term,} {to $-M_*$, i.e., minus the mass of the extremal solution in the absence of electric charge, defined in the right hand side of~\eqref{eq:masextnoQ}. The shift indicates that the mass and the electric charge are measured with respect to those of the extremal case with $Q_e=0=\mu$ which is considered here as a kind of ground state.} The Euclidean action is related to the Gibbs free energy ${\cal G}$ in the {following manner:
\begin{equation}
I_E=\beta {\cal G}=\beta\pqty{{\cal M}- T{\cal S}+ \Phi Q_e},
\end{equation}
where $\Phi\equiv \hat\sigma \mu$ is the electrostatic potential, identified with the chemical potential, $\mu$, of the dual theory. Now, from the boundary term (\ref{bterm}), it is easy to identify the various thermodynamic quantities. We read off
\begin{eqnarray}
{\cal M}=\hat \sigma(M-M_*),\quad {\cal S}=\mathcal S_o\pqty{1+\frac{\sqrt{2}\alpha p}{r_0}}\equiv \mathcal S_{\mathrm{W}}.
\label{ms}
\end{eqnarray}}

Finally, {we go back to the Lorentzian case by sending $Q_e\to iQ_e$, and we derive the Hawking temperature which reads
\begin{equation}
T=\frac{12 r_0^4-\pqty{2\hat{\eta} p^2+\mu^2} r_0^2}{8\pi r_0^2\left(2 r_0+\sqrt{2}\alpha p \right)}.\label{eq:Hawktemp}
\end{equation}
It is straightforward to verify that the first law, namely, $d{\cal M}=Td{\cal S}+\hat\sigma\mu dQ_e$, holds, if--f the slope $p$ of the axion profile is forbidden to fluctuate. The Hawking temperature is a monotonically increasing function of the horizon radius which vanishes at $r_0= r_*$, the latter defined in \eqref{eq:exthor} as the horizon radius of the extremal black hole. It will be also useful to have a formula for the event horizon radius --- the largest root of $F$ ---, $r_0$, in terms of the temperature. Using ~\eqref{eq:Hawktemp}, we can write
\begin{equation}
r_0=\frac{1}{6}\pqty{4\pi T+\sqrt{16\pi^2 T^2+24\sqrt{2}\pi \alpha p T+6\hat{\eta} p^2+3\mu^2}},\label{eq:r0temp}
\end{equation}
which will come in handy at a later stage.
}

\section{Holographic aspects\label{sec:Holo}}
Planar black holes with nontrivial axions distributed along the planar directions provide an ideal configuration for the computation of holographic DC responses. Their presence ensures that translation symmetry is broken which translates into momentum dissipation in the boundary theory; this simply opens the door to finite associated DC conductivities. In the boundary language, momentum relaxation simply means that $\nabla_t\ev{T^{tx^i}}\neq 0$. To see that this holds,  let us present a heuristic argument. Consider the four-dimensional renormalized/regularized version of the bulk action~\eqref{eq:toymodel}, and let us dub it $S_{\mathrm{ren}}$:
\begin{equation}
    S_{\mathrm{ren}}:=S_{\mathrm{bulk}}+S_{\mathrm{bdy}}+S_{\mathrm{ct}}.
\end{equation}
We saw that \eqref{eq:f4} is asymptotically AdS, hence it is conformally compact Einstein~\cite{Skenderis:2002wp}, and it can be brought to the Fefferman-Graham form
\begin{equation}
ds^2=\frac{d\rho^2}{4\rho^2}+\frac{\hat g_{\alpha\beta}(\rho,x)}{\rho}dx^{\alpha}dx^{\beta},
\end{equation}
the boundary being at $\rho=0$. We use Greek letters from the start of the alphabet to represent boundary indices, whereas $i,j,. . .$ are used for the spatial piece of the latter. The various fields admit the near-boundary expansions
\begin{eqnarray}
\hat{g}_{\alpha\beta}&=&\hat g^{(0)}_{\alpha\beta}+\rho g^{(2)}_{\alpha\beta}+. . .,\\
A_\alpha&=&A^{(0)}_{\alpha}+\rho^{1/2}A^{(1)}_{\alpha}+. . .,\\
\psi_i&=&\psi^{(0)}_{i}+\rho^{1/2}\psi^{(1)}_i+. . .,
\end{eqnarray}
Since $D=4$ there is no term in the series expansion related to the holographic conformal anomaly. For our purposes, determining the recursive relations between the coefficients is irrelevant, since we particularly care about the sources. For a euclidean boundary signature, the on-shell variation of $S_{\mathrm{ren}}$ yields
\begin{equation}
\var S_{\mathrm{ren}}=\int d^3x\sqrt{\hat{g}_0}\pqty{\frac{1}{2}\ev{T^{\alpha\beta}}\var \hat g^{(0)}_{\alpha\beta}+\ev{\mathcal{O}_i}\var \psi^{i}_{(0)}+\ev{\mathcal{J}^{\alpha}}\var A^{(0)}_\alpha},\label{eq:ref}
\end{equation}
where summation is implied for all repeated indices. Symmetry under the boundary $\mathrm{U}(1)$ transformation $\var A_\alpha^{(0)}=\nabla_\alpha\lambda$ implies $\nabla_\alpha\ev{\mathcal J^{\alpha}}=0$. Consequently, symmetry under boundary diffeomorphisms $\var \hat{g}_{\alpha\beta}^{(0)}=2\nabla_{(\alpha}\xi_{\beta)}$ leads to the diffeomorphism Ward identity
\begin{equation}
\nabla_\alpha\ev{T^{\alpha\beta}}=\ev{\mathcal{O}_i}\nabla^{\beta}\psi_{(0)}^{i}+F^{\beta}_{(0)\alpha}\ev{\mathcal{J}^\alpha}.
\end{equation}
Here, $\nabla_\alpha$ is the covariant derivative associated with the $\hat{g}_{(0)}$-compatible connection and $\xi^{\mu}=\{0,\xi^{\alpha}(x)\}$ is a boundary diffeomorphism-generating vector field, whereas $F^{(0)}_{\alpha\beta}=\nabla_\alpha A^{(0)}_\beta-\alpha\leftrightarrow \beta$. It is clear that since $\psi_{(0)}^i\sim  x^{i}$ by assumption, $\ev{T^{tx^i}}$ will not be conserved for $\ev{\mathcal{O}_i}\neq 0$. Hence, boundary momentum gets dissipated in the spatial directions, whereas the energy is of course conserved.

\subsection{Thermoelectric DC transport}
It was shown in~\cite{Donos:2014cya,Donos:2013eha} that the electric, thermoelectric and thermal conductivities can be computed in terms of the black hole horizon data alone without the need to invoke direct calculations on the (boundary) field theory side. This is achieved by properly manipulating the bulk field equations, revealing electric and heat currents which are manifestly independent of the holographic radial coordinate. These can be then evaluated at the horizon radius instead of the boundary. We start by considering the four-dimensional limit of \eqref{eq:toymodel}, an action functional of the metric $g_{\mu\nu}$, the gauge field $A_\mu$ and the two axion fields $\psi^1$ and $\psi^2$, where we take the bulk coordinates to be $x^{\mu}=\{t,r,x,y\}$. Studying the gauge field equations in the bulk, we observe that the only nonvanishing component is 
\begin{equation}
\partial_r(r^2F^{rt})=0.
\end{equation}
Defining the current density $\mathcal J^t=r^2 F^{tr}$, this corresponds to the charge density of the dual field theory when the right hand side is evaluated at the boundary, i.e. $Q\equiv \ev{\mathcal J^{t}}$, where $Q$ is the charge of the black hole, what will be $Q_e$ in our case. Moreover, we assume the existence of a regular horizon at $r_0$ (in the case of two horizons, the outer one is chosen), about which we assume the Taylor expansions $F\sim 4\pi T(r-r_0)+. . .$ and $\mathcal{A}\sim \mathcal{A}'(r_0)(r-r_0)+. . .$, namely we take the electric potential to vanish at the horizon radius. We will use Eddington-Finkelstein coordinates $(v,r)$ in order to make the regularity at the horizon apparent with $v=t+(4\pi T)^{-1}\ln(r-r_0)$. We will also assume the asymptotic behavior $\mathcal{A}\sim \mu-Qr^{-1}+. . .$ where $\mu$ is the chemical potential in the dual theory, defined as $\int_{r_0}^{\infty}dr F_{rt}$, while the dominant power in the asymptotic expansion of $F$ will be $\sim r^2$.

It is time to proceed with the perturbations. For starters, we will turn on a constant electric field of magnitude $E$ in the $x$ direction such that 
\begin{equation}
A_{x}=-\epsilon\bqty{Et- \mathcal{A}_{x}(r)},\label{eq:maxpert}
\end{equation}
supplemented by the small perturbations
\begin{equation}
g_{tx}=\epsilon h_{tx}(r),\qquad g_{rx}=\epsilon r^2 h_{rx}(r),\qquad \psi^1=px+\epsilon\mathcal{X}(r),\label{eq:metpert}
\end{equation}
about the black hole background given by \eqref{eq:f4} and $\mathcal{A}=\mu-Q_e/r=\mu(1-r_0/r)$. Here, $\epsilon$ is introduced as a small parameter helping us keep track of the perturbation order. We will now study the gauge field current density which possesses only one nontrivial component, the one in the $x$ direction,
\begin{equation}
\mathcal{J}^{x}=-\pqty{F\mathcal{A}'_x+\frac{Q_e h_{tx}}{r^2}}.\label{eq:curr}
\end{equation}
This can be evaluated at any $r$, and it is radially conserved since it is derived by integrating the equation $\partial_r(\sqrt{-g}F^{xr})=0$. This means that we are allowed to evaluate it at the horizon radius instead of the boundary. 

Next, we look at the metric field equations. We observe that $\mathcal{G}_{rx}=0$ is an equation algebraic in $h_{rx}$ which is solved by
\begin{equation}
h_{rx}=\frac{\mathcal{X}'}{p}-\frac{2EQ_e}{p r^2 F(2\hat{\eta} p+\sqrt{2}\alpha F')},\label{eq:hrx}
\end{equation}
where $F$ is always on the background shell since we used the fact that $\mathcal{G}_{yy}=0$ to arrive at this particular expression. The linearized axion field equations also follow from \eqref{eq:hrx}. In addition, we also have the second order inhomogeneous ordinary differential equation (ODE):
\begin{equation}
{r(\sqrt{2}\alpha p+2 r)F}h''_{tx}-\sqrt{2}\alpha p Fh'_{tx}-\bqty{4 F+p\pqty{2\hat{\eta} p+\sqrt{2}\alpha F'}}h_{tx}+2Q_eF\mathcal{A}'_x=0,\label{eq:htx}
\end{equation}
which corresponds to $\mathcal{G}_{tx}=0$. To move on, we need to impose boundary conditions. 

We first need to check the gauge field perturbation and its regularity at the horizon. In Eddington-Finkelstein coordinates, the full gauge field perturbation reads
\begin{equation}
A_{x}=-\epsilon\pqty{Ev-\mathcal{A}_x+\frac{E\ln(r-r_0)}{4\pi T}}.
\end{equation}
Taylor-expanding this about $r_0$ one sees that its regularity is ensured only if 
\begin{equation}
\mathcal A_x=-\frac{E\ln(r-r_0)}{4\pi T}+\mathcal{O}(r-r_0).
\end{equation}
It is also evident that near the horizon
\begin{equation}
\mathcal{A}'_x\sim -\frac{E}{4\pi T(r-r_0)}+. . .\sim -\frac{E}{F}+. . .,
\end{equation}
because $4\pi T=F'(r_0)$ and $F\sim F'(r_0)(r-r_0)+. . .$. Now, we can also see that \eqref{eq:hrx} diverges as $r\to r_0$ because of the presence of $F$ in the denominator. In order to cure this, we let $h_{tx}$ expand as
\begin{equation}
h_{tx}=-\frac{2EQ_e}{p (2\hat{\eta} p+\sqrt{2}\alpha F')}+\mathcal{O}(r-r_0).
\end{equation}
near the horizon. Then, one can immediately see that \eqref{eq:htx} vanishes when evaluated at $r_0$. As for the axion field perturbation $\mathcal{X}$, we just assume a constant value at $r_0$ and sufficient falloff at infinity. The remaining boundary conditions at radial infinity are discussed in~\cite{Donos:2014cya} in detail. Having established well posed perturbations of the bulk fields, we can easily extract the electric DC conductivity, by first evaluating \eqref{eq:curr} at $r_0$ and at leading order in $(r-r_0)$, further dividing by the external electric field of magnitude $E$, i.e.,  $\sigma=\ev{\mathcal{J}^{x}}/E$ at the horizon. We find that 
\begin{equation}
\sigma=1+\frac{Q_e^{2}}{(\hat{\eta}p+2\sqrt{2}\pi\alpha T)pr_0^2}=1+\frac{\mu^2}{\hat{\eta} p^2+2\sqrt{2}\pi\alpha p T}.\label{eq:elcondu}
\end{equation}
This is in perfect agreement with~\cite{Andrade:2013gsa} when $\alpha=0$. 

The nonminimal coupling of the axion fields to gravity modifies the electric conductivity compared to the results obtained in the case of the EMA theory. However, and most importantly, the behavior at both temperature extremes is the same. As we saw when we studied the horizon structure, zero temperature corresponds to $r_*$ which is independent of $\alpha$, and thus matches the horizon radius of the extremal EMA solution. The electric conductivity at $T=0$ is obtained by the replacement $r_0\to r_*$ in~\eqref{eq:elcondu}. It is finite, and it obviously agrees with the result in~\cite{Andrade:2013gsa}, whereas when $T\to \infty$, $\sigma$ goes to unity which is again the standard conducting behavior extracted from an EMA bulk. Such a behavior has also been observed in the pertinent cases~\cite{Feng:2015oea,Baggioli:2017ojd}. Noticeably, the result~\eqref{eq:elcondu} satisfies the $\sigma\geq 1$ bound proposed in~\cite{Grozdanov:2015qia}, regardless of the dissipation strength. It is clear that since the new coupling does not enter into the leading order of the expansions about the two temperature extremes, one cannot expect deviations. To continue, we need to consider a time-dependent source for the heat current in our perturbation ansatz. This will allow us to compute the thermoelectric conductivities, $\upalpha$, $\bar{\upalpha}$ and the thermal conductivity $\bar{\kappa}$ at zero electric field, thus filling the remaining entries of the transport matrix. 

We consider the ans\"atze \eqref{eq:maxpert} and \eqref{eq:metpert}, but now we switch on a time-dependent part in $g_{tx}$, namely $g_{tx}=\epsilon(tf_2(r)+h_{tx}(r))$, while we make a more general ansatz for the gauge field; in particular, $A_x=\epsilon(tf_1(r)+\mathcal{A}_x(r))$. The $x$ component of the gauge equations of motion is neatly written as a radial conservation law for the only nonvanishing component of the current density in the spatial directions, $\mathcal{J}^{x}=r^2F^{xr}$,
\begin{equation}
\mathcal{J}^{x}=-\bqty{F\mathcal{A}'_x+\frac{Q_e h_{tx}}{r^{2}}+t\pqty{F f'_1-\frac{Q_ef_2}{r^2}}}.\label{eq:currtherm}
\end{equation}
Using the radial conservation of \eqref{eq:currtherm} together with the unperturbed field equations, we can manage to find a first $r$-integral of $-2\mathcal{G}_{tx}$, namely, the radially constant quantity
\begin{equation}
\mathcal{Q}^{x}=\pqty{1+\frac{\alpha p}{\sqrt{2}r}}F^2\pqty{\frac{g_{tx}}{\epsilon F}}'-\mathcal{A}\mathcal{J}^{x},\label{eq:heatcur}
\end{equation}
which we can identify with the $x$ component of the heat current of the boundary theory when evaluated at $r\to \infty$. Again, since this is radially conserved, we can evaluate it at $r_0$ instead. Additionally, the $rx$ component of the metric field equations is an algebraic equation for the relevant perturbation which is solved by 
\begin{equation}
h_{rx}=\frac{\mathcal{X}'}{p}+\frac{(2r+\sqrt{2}\alpha p)(rf_2'-2f_2)+2Q_ef_1}{{(2\hat{\eta} p^2+\sqrt{2}\alpha p F')Fr^2}}.
\end{equation}
Indeed, as a consistency check, killing $f_2$ and setting $f_1=-E$ yields \eqref{eq:hrx} as it should. 

We see that if we choose $f_2=-\gamma F$ and $f_1=\gamma \mathcal{A}-E$, all time dependence vanishes in $\mathcal{J}^{x}$ and in the $tx$ component of the metric field equations, the former assuming the expression \eqref{eq:curr}, while the latter becoming \eqref{eq:htx}. In the Eddington-Finkelstein coordinate system, the regularity of the bulk perturbations and the satisfaction of the perturbed field equation $\mathcal{G}_{tx}=0$ near the horizon radius, both boil down to the series expansion
\begin{equation}
h_{tx}\sim-\frac{EQ_e+2\gamma \pi r_0(2r_0+\sqrt{2}\alpha p) T}{\hat{\eta} p^2+2\sqrt{2}\pi \alpha p T}-\frac{\gamma F\ln(r-r_0)}{4\pi T}+. . .,
\end{equation} 
which in turn leads to the radially-constant quantities 
\begin{eqnarray}
\ev{\mathcal{J}^{x}}&=&E\sigma+\gamma\frac{2\mu\pi r_0(\mathcal{S}+\mathcal{S}_o) T}{\mathcal{S}_o(\hat{\eta} p^2+2\sqrt{2}\pi \alpha p T)},\\
\ev{\mathcal{Q}^{x}}&=&E\partial_\gamma\!\ev{\mathcal{J}^{x}}_{r_0}+\gamma\frac{\pi\pqty{\mathcal{S}+\mathcal{S}_o}^2T^2}{\mathcal{S}_o(\hat{\eta} p^2+2\sqrt{2}\pi \alpha p T)},
\end{eqnarray}
where $\mathcal{S}$ is defined in \eqref{ms}, $\mathcal{S}_o:=4\hat \sigma \pi r_0^2$ and $\sigma$ is as in~\eqref{eq:elcondu}. Clearly, we have all the necessary transport coefficients of the strongly coupled theory, and we can now explicitly write down the generalized Ohm/Fourier law,
\begin{equation}
\pmqty{\ev{\mathcal{J}^x}\\\ev{\mathcal{Q}^x}}=\pmqty{\sigma&{\upalpha} T\\\bar{\upalpha}T&\bar{\kappa}T}\pmqty{E\\-\nabla_x T/T},\label{eq:transmat}
\end{equation}
from which we can read off the linear DC reponse of the system to an external electric field and a thermal gradient. Here, 
\begin{equation}
{\upalpha}=\frac{\partial_\gamma\!\ev{\mathcal{J}^{x}}}{T}=\frac{2\mu\pi r_0(\mathcal{S}+\mathcal{S}_o) }{\mathcal{S}_o(\hat{\eta} p^2+2\sqrt{2}\pi \alpha p T)},\quad \bar{\upalpha}=\frac{\partial_E\!\ev{\mathcal{Q}^{x}}}{T}={\upalpha},\quad \bar{\kappa}=\frac{\pi\pqty{\mathcal{S}+\mathcal{S}_o}^2T}{\mathcal{S}_o(\hat{\eta} p^2+2\sqrt{2}\pi \alpha p T)},
\end{equation}
are the thermoelectric conductivities and the thermal conductivity at zero electric field, respectively. First of all, the transport matrix \eqref{eq:transmat} is symmetric which constitutes a successful consistency check against the Onsager relations~\cite{Donos:2017mhp} for theories invariant under time reversal, the latter relating the current densities of the background geometry to their counterparts obtained from a time-reversed solution. Secondly, when $\alpha\to 0$, the coefficients successfully reduce to those of EMA theory obtained as a special example in~\cite{Donos:2014cya}.

\subsection{Bounds for thermal conductivity and diffusion constants}
In this subsection we wish to probe the theory against various relevant bounds in the holography-related literature. With the complete set of conductivities at hand we can work out some interesting relations from~\eqref{eq:transmat}. First of all, let $\mathcal{J}\equiv \ev{\mathcal{J}^{x}}$ and $\mathcal{Q}\equiv \ev{\mathcal{Q}^{x}}$. We have that
\begin{equation}
\pqty{{\frac{\mathcal{J}}{E}}}_{\mathcal{Q}=0}=\sigma-\frac{{\upalpha}^2T}{\bar{\kappa}}=1,
\end{equation}
which ultimately represents the conductivity in the absence of heat flows. In addition, the simple relation discussed in~\cite{Donos:2014cya} is modified:\footnote{We set $\hat \sigma=1$ in the entropy expression from now on}
\begin{equation}
\frac{\bar{\kappa}}{{\upalpha}}=\frac{(\mathcal{S}+\mathcal{S}_o)T}{2Q_e}=\frac{\mathcal{S}_oT}{Q_e}+\frac{2\sqrt{2}\pi \alpha p T}{\mu}.\label{eq:ka}
\end{equation}
From the transport matrix we can also define the thermal conductivity at zero electric current as 
\begin{equation}
\kappa=\bar{\kappa}-\frac{{\upalpha}^2T}{\sigma}=\frac{\bar{\kappa}}{\sigma}=\frac{\pi(\mathcal{S}+\mathcal{S}_o)^2T}{\mathcal{S}_o\pqty{\mu^2+\hat{\eta} p^2+2\sqrt{2}\pi \alpha p T}}.
\end{equation}
Moreover, the Lorentz ratios 
\begin{equation}
\bar L=\frac{\bar{\kappa}}{\sigma T}=\frac{\kappa}{T},\qquad L=\frac{\bar{L}}{\sigma}=\frac{\pi(\mathcal{S}+\mathcal{S}_o)^2(\hat{\eta} p^2 +2\sqrt{2}\pi \alpha p T)}{\mathcal{S}_o\pqty{\mu^2+\hat{\eta} p^2+2\sqrt{2}\pi \alpha p T}^2},
\end{equation}
will be of interest as well. We observe that $\sigma$, ${\upalpha}$ and $\bar{\kappa}$ blow up as $p\to 0$, whereas ${\kappa}$ goes to the finite value $4\pi \mathcal{S}_oT/\mu^2$, $\bar{L}\to 4\pi \mathcal{S}_o/\mu^2$ and $L\to 0$. Moreover, at zero temperature, we notice an electric conductor/thermal insulator behavior which is reminiscent of the findings in the much simpler linear axion model~\cite{Andrade:2013gsa}. 

\paragraph{Violation of the thermal conductivity bound.} {Positivity of the temperature suggests that 
\begin{equation}
\mathcal{S}_o=16\pi G_{\mathrm{eff}}\mathcal{S}\geq \frac{\pi(2\hat{\eta} p^2+\mu^2)}{3},
\end{equation}
where $G_{\mathrm{eff}}$ is defined in~\eqref{eq:kappaef}. Thus, since $\mathcal{S}\geq \mathcal{S}_o$, the inequality being saturated for $\alpha=0$, we find that 
\begin{equation}
\kappa\geq  \frac{4\pi^2(\hat{\eta} p^2+\mu^2)T}{3\pqty{\mu^2+\hat{\eta} p^2+2\sqrt{2}\pi \alpha p T}}.\label{eq:kTbound}
\end{equation}
From this, we can extract a lower bound, which we shall contrast with the universal bound proposed in~\cite{Grozdanov:2015djs}, i.e., $\kappa/T\geq 4\pi^2/3$. To do so, we reformulate~\eqref{eq:kTbound} using the ratios $\tilde{p}:=p/T$ and $\tilde{\mu}:=\mu/T$:
\begin{equation}
\bar{L}\equiv\frac{\kappa}{T}\geq \frac{4\pi^2(\hat{\eta} \tilde p^2+\tilde \mu^2)}{3\pqty{\tilde \mu^2+\hat{\eta} \tilde p^2+2\sqrt{2}\pi \alpha \tilde p}}=:\mathcal{B}_{\kappa}.
\end{equation}
Taylor-expanding $\mathcal{B}_\kappa$ about small and large $\tilde{p}$ yields two series with a leading order equal to $4\pi^2/3$. However, the expansions about small and large $\tilde{\mu}$, 
\begin{equation}
\mathcal{B}_{\kappa}=\frac{4\pi^2}{3} \frac{\tilde p}{\tilde p+2\sqrt{2}\alpha \pi/\hat\eta}+\mathcal{O}(\tilde{\mu}^2),\qquad \mathcal{B}_{\kappa}=\frac{4\pi^2}{3}+\mathcal{O}(1/\tilde{\mu}^2),\label{eq:largemu}
\end{equation}
respectively, give rise to a more interesting picture: since
\begin{equation}
   \frac{\tilde p}{\tilde p+2\sqrt{2}\alpha \pi/\hat\eta}<1,
\end{equation}
there is a parametric violation of the $4\pi^2/3$ bound, provided that the dissipation scale does not dominate over $\alpha/\hat\eta$. It is perhaps more fitting to write $\mathcal{B}_\kappa$ in terms of the variable $\Breve{p}:=\tilde{p}\sqrt{\hat\eta}/\tilde{\mu}$ and the rescaled coupling $\Breve{\alpha}:=\alpha/(\tilde{\mu}\sqrt{\hat\eta})$:
\begin{equation}
    \mathcal{B}_\kappa=\frac{4\pi^2}{3}\frac{1+\Breve{p}^2}{1+\Breve{p}^2+2\sqrt{2}\pi \Breve{\alpha}\Breve{p}}.
\end{equation}
This exhibits a global minimum at $\Breve{p}=1$, or equivalently, at ${p}\sqrt{\hat\eta}={\mu}$, which reads
\begin{equation}
\min \mathcal{B}_\kappa=\frac{4\pi^2}{3}\frac{1}{1+\sqrt{2}\pi\Breve{\alpha}}.\label{eq:kmin}
\end{equation}
Since $\Breve{\alpha}$ is positive, this is always less than $4\pi^2/3$. Ergo, the proposed bound is certainly violated, but it is not replaced by another constant function, i.e., a lower fixed number, valid at all scales. To the contrary, the violation --- merely an artefact of the particular gravity bulk deformation introduced in the beginning of this work --- is parametric, with the bounding function rather depending on $\Breve{\alpha}$ and $\Breve{p}$. }
\begin{figure}[ht!]
\centering
\includegraphics[width=0.7\textwidth]{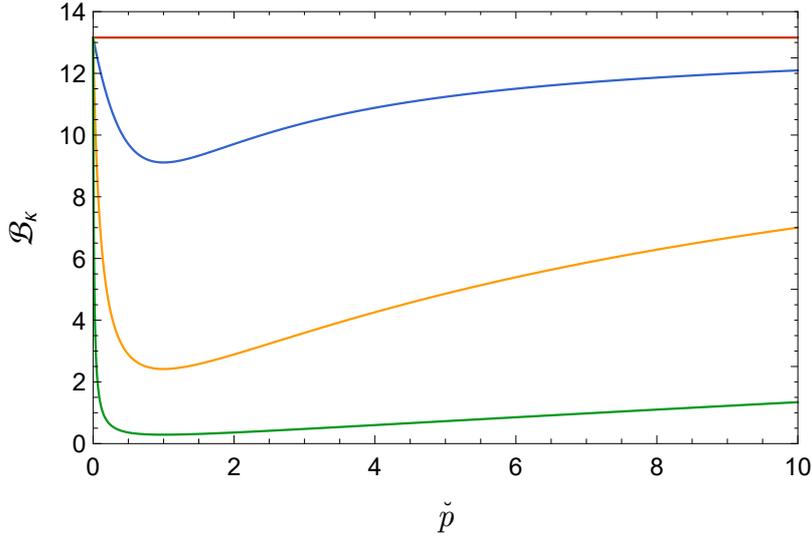}
\caption{The red line always indicates the $4\pi^2/3$ bound when $\Breve\alpha=0=\alpha$. The blue line is for $\Breve{\alpha}=0.1$, the yellow for $\Breve{\alpha}=1$, and the green for $\Breve{\alpha}=10$.}
\label{fig:kT}
\end{figure}
\newline
{From figure~\ref{fig:kT} we can see that for a given chemical potential at fixed temperature with $\mu\sqrt{\hat\eta}\ll \alpha T$, the heat conductivity bound can be significantly lowered, with the lowering becoming more effective close to dissipation strengths of order $p\sim \mu/\sqrt{\hat \eta}$. However, as expected from~\eqref{eq:kmin}, the bounding function cannot be driven all the way down to zero. Finally, the upper bound,  proposed in~\cite{Donos:2014cya}, still holds in terms of the modified entropy, i.e., \begin{equation}
\bar{L}\leq \frac{4\pi\mathcal{S}^2}{\mathcal{S}_o\mu^2}=\frac{\mathcal{S}^2}{Q_e^2}.
\end{equation}}

\paragraph{The Kelvin formula.} Another celebrated relation, the Kelvin formula, attributed to holographic models flowing towards an $\text{AdS}_2\times \mathds R^{2}$ fixed point in the IR~\cite{Blake:2016jnn,Davison:2016ngz}, reads 
\begin{equation}
\pqty{\frac{\upalpha}{\sigma}}_{T=0}\equiv \lim\limits_{T\to 0}\pqty{\pdv{\mathcal{S}}{Q}}_{T},\label{eq:KelvinForm}
\end{equation}
where $Q\equiv Q_e$ is the charge density in our case. Motivated by the presence of such a near-horizon geometry in our case study, we wish to probe the validity of~\eqref{eq:KelvinForm}. First of all, the Seebeck coefficient,  $\upalpha/\sigma$, at zero temperature is
\begin{equation}
\pqty{\frac{\upalpha}{\sigma}}_{T=0}=\frac{2\pi\mu\pqty{3\sqrt{2}\alpha p+\sqrt{6\hat{\eta} p^2+3\mu^2}}}{3(\mu^2+\hat{\eta} p^2)}.\label{eq:seebeck}
\end{equation}
Then, we can use the chain rule in order to write 
\begin{equation}
\pdv{\mathcal{S}}{Q_e}=\pdv{\mathcal{S}}{\mu}\pqty{\pdv{Q_e}{\mu}}^{-1}.\label{eq:KelvinRHS}
\end{equation}
Using the handy relation~\eqref{eq:r0temp} together with $Q_e=\mu r_0$, and taking the $T\to 0$ limit of~\eqref{eq:KelvinRHS} afterwards, we indeed arrive at~\eqref{eq:seebeck}, proving its validity. Next, we wish to investigate the thermoelectric response in the diffusion-dominated regime. 

\paragraph{Lower bounds for the diffusion constants.} The incoherent limit is defined by $p\gg T,\mu$ for fixed $\tilde{\mu}$. When dissipation dominates, the transport coefficients expand as 
\begin{equation}
\sigma=1+\mathcal{O}(1/p^2),\quad \upalpha=\frac{2\sqrt{2}\pi\alpha\mu}{\hat{\eta} p}+\mathcal{O}(1/p^2),\quad \bar\kappa=\frac{8\pi^2(\sqrt{\hat{\eta}}+\sqrt{3}\alpha)^2}{3\hat{\eta}}+\mathcal{O}(1/p),
\end{equation}
with $\kappa$ having the same leading order coefficient as $\bar{\kappa}$. In this regime, diffusion takes over, and the horizon radius goes as $\propto p$, in particular, $r_0=p\sqrt{\hat{\eta}/6}$. One can observe that the off-diagonal elements of the transport matrix~\eqref{eq:transmat} have an $\mathcal{O}(1/p)$-falloff for large $p$, whereas the diagonal ones go to a finite value. The ratio of charged to neutral degrees of freedom measured by the $Q_e/\mathcal{S}$ ratio goes as $\sim \mu/(\alpha p)\to 0$, and the charge/heat currents decouple~\cite{Davison:2015bea}. A priori, we will not assume that the charge and energy diffusitivities decouple, namely, we will not neglect the mixing term, $\mathfrak{M}$, defined below. The coupled diffusion is  described~\cite{Kim:2017dgz} by the constants 
\begin{equation}
D_{\pm}=\frac{a_1\pm \sqrt{a_1^2-4a_2}}{2},\label{eq:diffu}
\end{equation}
with 
\begin{equation}
a_1:=\frac{\sigma}{\chi}+\frac{\kappa}{c_{Q_e}}+\mathfrak{M},\quad a_2:=\frac{\sigma\kappa}{\chi c_{Q_e}},\quad \mathfrak{M}:=\frac{(\zeta\sigma-\upalpha \chi)^2T}{\sigma c_{Q_e} \chi^2},
\end{equation}
where $\chi$, $\zeta$ and $c_{Q_e}$ are the charge susceptibility, thermoelectric susceptibility and specific heat at fixed charge density $Q_e$, respectively. In whatever regime $\zeta,\upalpha=0$, the diffusitivities do, in fact, decouple with $D_+\to D_c:=\sigma/\chi$ (charge diffusion constant) and $D_-\to D_e:=\kappa/c_{Q_e}$ (energy diffusion constant).

\begin{figure}[ht!]
\centering
\subfigure[$\tilde{\mu}=0.01$]{
\includegraphics[width=.45\textwidth]{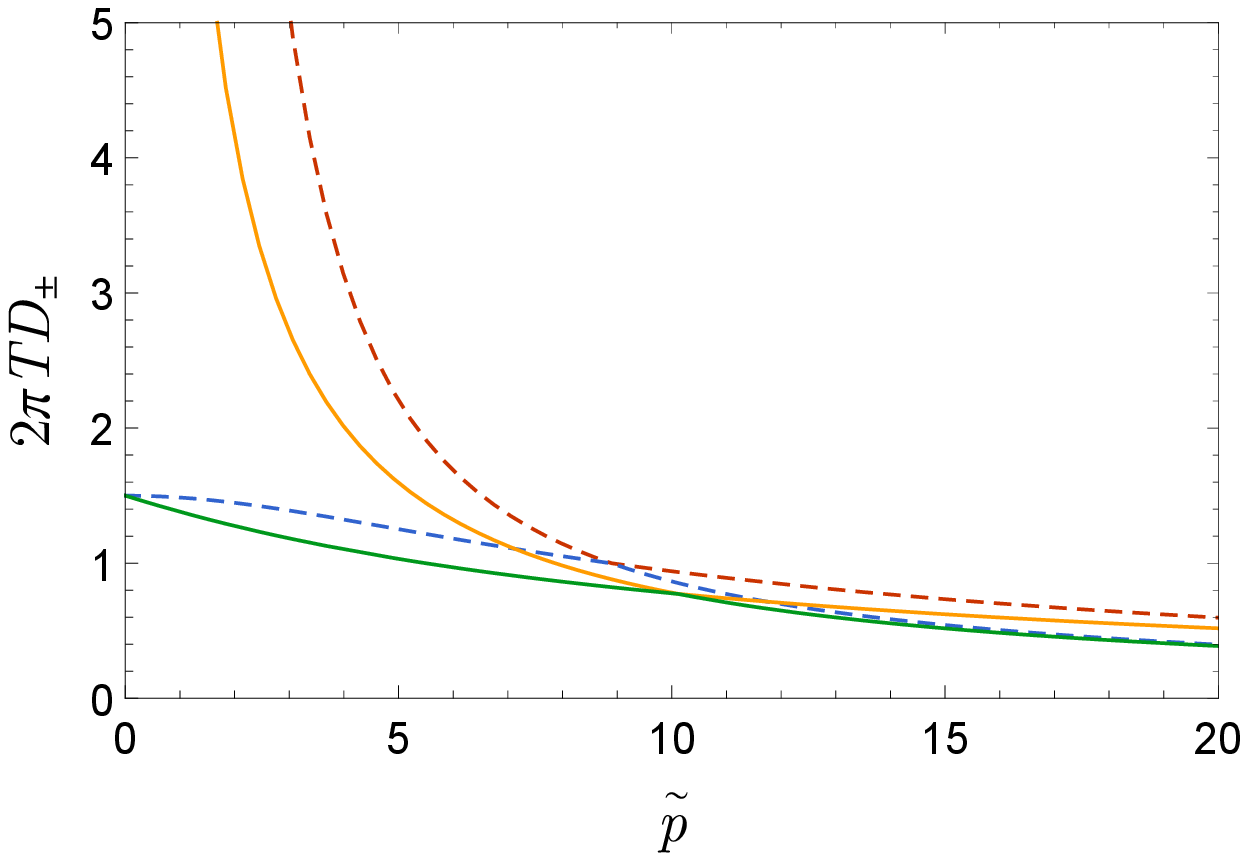}
}
\subfigure[$\tilde{\mu}=10$]{
\includegraphics[width=.45\textwidth]{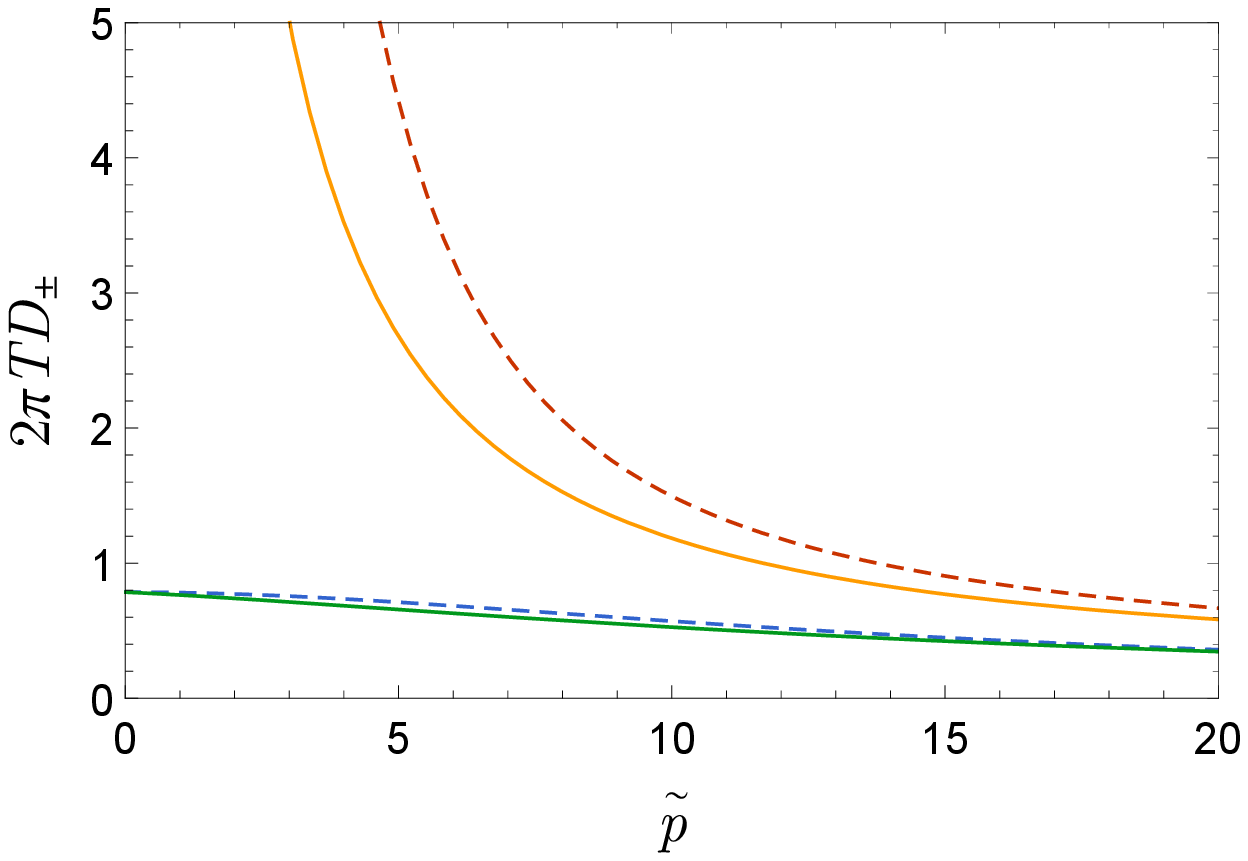}
}

\subfigure[$\tilde{\mu}=0.01$ and $\alpha=0.1$]{
\includegraphics[width=.45\textwidth]{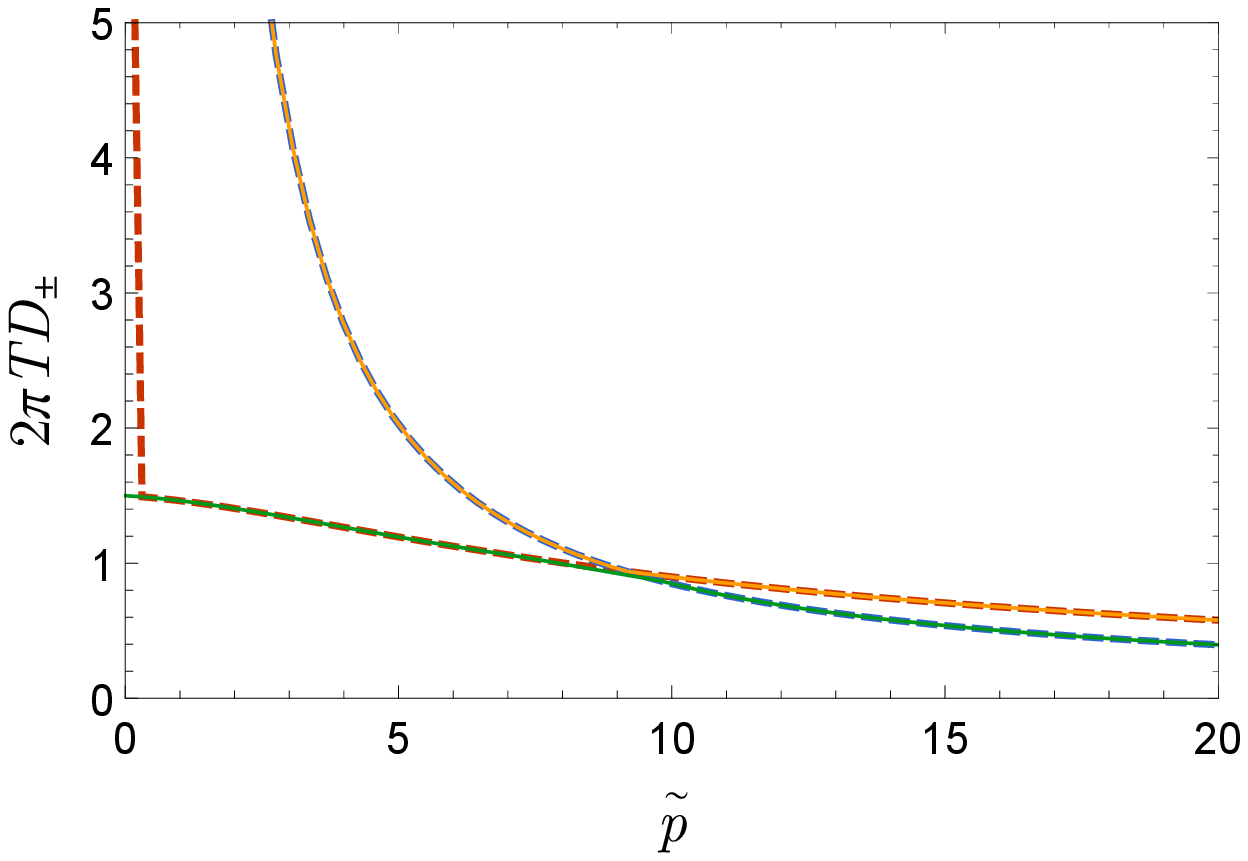}
}
\subfigure[$\tilde{\mu}=10$ and $\alpha=0.11$]{
\includegraphics[width=.45\textwidth]{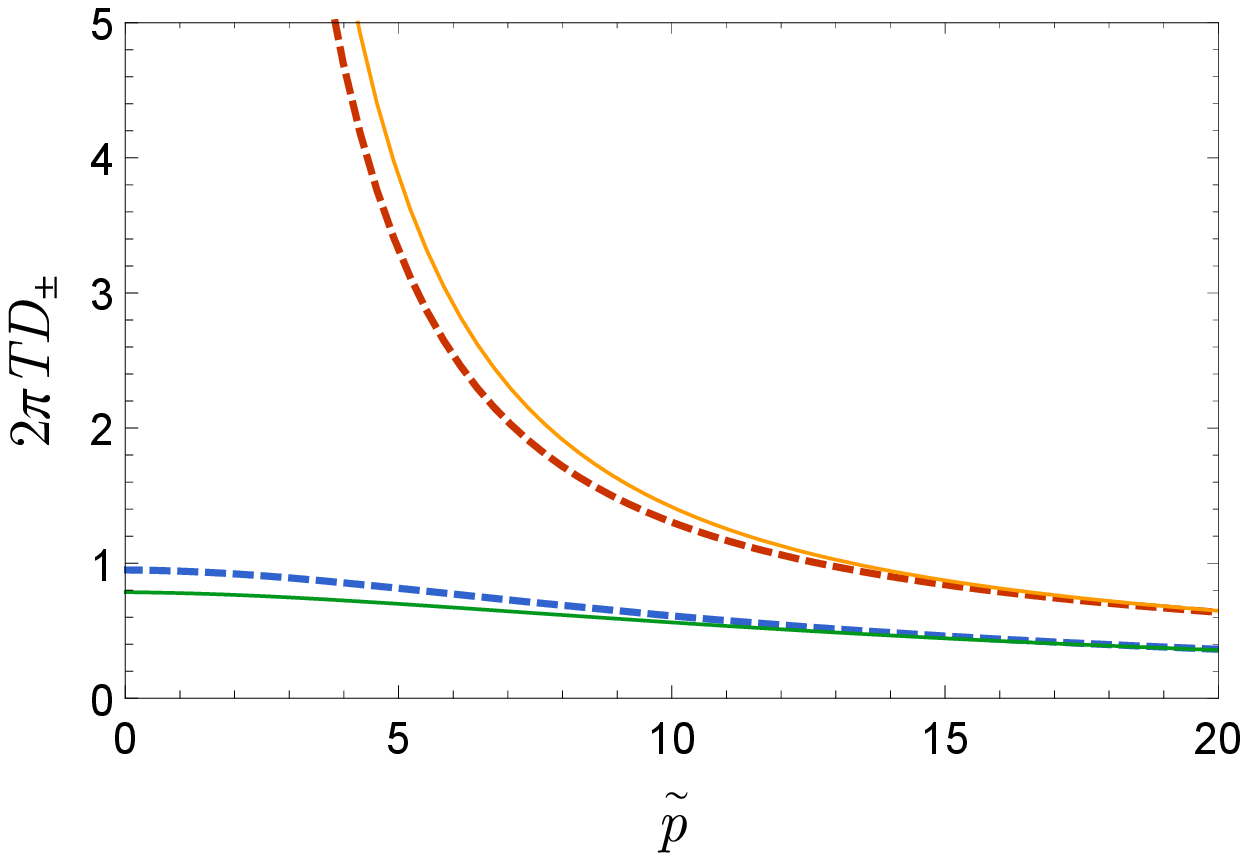}
}

\caption{We considered $\hat\eta=1$ for the plots. In subfigures (a) and (b), the dashed lines are for $\alpha=0$, whereas the solid ones are for $\alpha=0.5$. Red and yellow lines show $2\pi TD_+$ vs. $\tilde{p}$, whereas blue and green ones depict $2\pi TD_-$ vs. $\tilde{p}$. In subfigures (c) and (d), we display the mixing feature. The red (blue) thick dashed line shows $2\pi TD_{c(e)}$ vs. $\tilde{p}$, whereas the yellow (green) solid one depicts $2\pi TD_{+(-)}$ vs. $\tilde{p}$.}
\label{fig:diffu}
\end{figure}

{Let us first compute the thermodynamic susceptibilities. We have 
\begin{eqnarray}
\chi&:=&\pqty{\pdv{Q_e}{\mu}}_T=\frac{1}{6}\pqty{4\pi T +\frac{{{C}}^2+3\mu^2}{{{{C}}}}},\\
\zeta&:=&\pqty{\pdv{\mathcal{S}}{\mu}}_T=\frac{2\pi\mu}{3}\pqty{1+\frac{4\pi T+3\sqrt{2}\alpha p}{{{{C}}}}},
\end{eqnarray}
where ${{C}}:=\sqrt{16\pi^2 T^2+24\sqrt{2}\pi\alpha p T+6\hat{\eta} p^2+3\mu^2}\equiv 6r_0-4\pi T$ for safety of space. The specific heat at fixed $Q_e$ is given by 
\begin{equation}
c_{Q_e}:=c_\mu-\frac{\zeta^2 T}{\chi}=\frac{({{C}}+4\pi T)c_\mu}{6\chi},
\end{equation}
where
\begin{equation}
c_\mu:=T\pqty{\pdv{\mathcal{S}}{T}}_\mu=2CT\pqty{\frac{\zeta}{\mu}}^2,
\end{equation}
is the specific heat at fixed chemical potential. Plugging everything back into~\eqref{eq:diffu}, we can get an explicit expression for the diffusitivities at all $\tilde{p}$ scales. The explicit expressions are too lengthy and not so enlightening, hence we simply plot the results in figure ~\ref{fig:diffu}. }

{From the last-mentioned figure, subfigures (c) and (d) in particular, there are two observations to be made. Clearly, as $\tilde{\mu}$ increases, the mixing term has a decreasing impact and the charge/energy diffusitivities  decouple. Indeed, $\mathfrak M$ falls off as $\mathcal{O}(1/\tilde{\mu})$. On the other hand, for small $\tilde{\mu}$, subfigure (c) shows that the mixing term becomes maximal, leading to a completely opposite identification of $D_{\pm}$ with $D_c$ and $D_e$, opposite to the way these quantities are matched in the incoherent phase; for $\tilde{\mu}\ll 1$, $D_+$ is identified with $D_e$, whereas $D_-$ is identified with $D_c$. Figure~\ref{fig:diffu} also makes apparent that, regardless of the value of $\tilde{\mu}$, the mixed diffusion constants do completely decouple in the incoherent limit. Subfigure (c) proves to be very indicative of this fact. One can locate the decoupling at $\tilde{p}\sim 10$ which means that the system is described in terms of the charge/energy diffusitivities for dissipation strengths $p \gtrsim 10 T\gg T$
and $p\gtrsim 10^3\mu\gg \mu$. Since $p\gg T,\mu$, this certainly lies in the diffusion-dominated region. Hence, it is safe to say that $D_+\to D_c$ and $D_-\to D_e$ when dissipation becomes strong. Indeed, the large-$p$ expansions of $D_\pm$ read 
\begin{equation}
TD_+=\frac{\sqrt{6}}{\sqrt{\hat{\eta}}\tilde p}+\mathcal{O}(1/\tilde p^2),\quad TD_-=\frac{\sqrt{3}}{\sqrt{2\hat{\eta}}\tilde p}+\mathcal{O}(1/\tilde p^3),\label{eq:diffuexp}
\end{equation}
exhibiting a leading order agreement with the expansions of $D_c$ and $D_e$, respectively, in the incoherent limit. We mention here that these results have also been derived in~\cite{Kim:2017dgz}. Apart from a mild curve shifting displayed in figure~\ref{fig:diffu}, the new coupling has no effect at the dissipation extremes; it does not contribute to the leading order in~\eqref{eq:diffuexp}, whereas it has no qualitative effect at small $\tilde{\mu}$, where the mixing becomes maximal.}

After all this song and dance, the ultimate aim is to see if the new coupling affects the diffusitivity bound proposal in~\cite{Blake:2016wvh,Blake:2016sud}. According to the Hartnoll conjecture~\cite{Hartnoll:2014lpa}
\begin{equation}
D\gtrsim v^2\frac{\hbar}{k_BT},
\end{equation}
for $v$ being some characteristic velocity. Instead of the original idea to match the latter with the speed of light, a reasonably natural candidate for $v$ at strong coupling has been the butterfly velocity, a measure of the spatial propagation speed of chaos through the dual quantum system. This has been derived in~\cite{Blake:2016wvh} for a general IR geometry
\begin{equation}
ds^2_D=-F(r)dt^2+\frac{dr^2}{F(r)}+V(r)dx^{i}dx_i,\quad i=1,\dots,D-2,
\end{equation}
with matter, minimally coupled to an Einstein gravity bulk. The holographic derivation depends on the black hole horizon data, and its geometric picture is that of a shock wave propagating in the bulk; the butterfly effect is manifest through the late-times exponential boosting of the energy of an in-falling particle near the black hole horizon. For more details, please see~\cite{Shenker:2013pqa,Roberts:2014isa}. {We are interested in the --- expected to be --- numbers $\mathcal{B}_{c(e)}$ which act as a lower bound in the incoherent phase, i.e., 
\begin{equation}
\frac{2\pi T D_{c(e)}}{v_B^2}\geq \mathcal{B}_{c(e)}.
\end{equation}
So, in order to determine these, we need to compute the butterfly velocity first.}

{A lightning quick calculation with Mathematica using standard methods in~\cite{Blake:2016wvh,Shenker:2013pqa,Roberts:2014isa}, and the more relevant~\cite{Kim:2017dgz} in particular, reveals that the screening length $m$ is not modified and the $uu$ component of the perturbed equations at linearized order reads 
\begin{equation}
(\partial_i\partial_i-m^2)h(x,t_w)\sim f(A(0),V(0),\alpha)E_0e^{2\pi t_w/\beta}\delta(x),\label{eq:shift}
\end{equation}
with 
\begin{equation}
m^2\sim \pqty{\frac{\partial_{uv}V(uv)}{A(0)}}_{uv=0}.
\end{equation}
Here, the expression is in Kruskal coordinates $(u,v)$ with $A,V$ functions of $uv$, $f$ is some --- irrelevant to the solution --- function with $\alpha$ being part of its arguments, $\beta=1/T$, $t_w$ is the past time the particle was released on the boundary of AdS,\footnote{The expression is valid for late times $t_w$ greater than the thermal timescale $\beta$.} and $E_0$ is the initial energy of that particle. Since the screening length is not modified, there will be no deviation, either in the value of the Lyapunov exponent $\lambda_L$, or in the expression for the butterfly velocity $v_B$. The solution to~\eqref{eq:shift} has essentially the same form as if the gravity bulk was pure Einstein gravity. Thus, comparing it with the exponential formula determining the growth in the commutators of generic Hermitian local operators, e.g., see relation (4) in~\cite{Blake:2016wvh}, we deduce that $\lambda_L=2\pi/\beta$ and \begin{equation}
v_B^2=\frac{4\pi^2}{(\beta m)^2 }=\frac{\pi T}{r_0}=\frac{6\pi}{4\pi+\sqrt{16\pi^2+24\sqrt{2}\pi \alpha \tilde p+ 6\hat{\eta}\tilde p^2+3\tilde{\mu}^2}}.
\end{equation}
In the incoherent limit, the butterfly velocity squared goes to 0 as $\sim 1/\tilde p$, and thus, since $\alpha$ does also not contribute to the leading order of the expansions~\eqref{eq:diffuexp}, we expect that the specific gravity bulk deformation will not affect the universal diffusitivity bounds. It is important to stress that these bounds are universal only in the diffusion-dominated regime. Indeed, expanding the ratios about large $\tilde{p}$, we find that
\begin{align}
    \frac{2\pi TD_c}{v_B^2}&=2+\frac{3\tilde{\mu}^2-16\pi^2}{3\hat\eta\tilde{p}^2}+\mathcal{O}(1/\tilde{p}^3),\\ \frac{2\pi TD_e}{v_B^2}&=1+\frac{2\sqrt{2}\pi\pqty{3\alpha+\sqrt{3\hat\eta}}}{2\hat\eta\tilde{p}}+\mathcal{O}(1/\tilde{p}^2),
\end{align}
and one can verify that 
\begin{equation}
    \frac{2\pi TD_c}{v_B^2}\geq 2=:\mathcal{B}_c \qquad \frac{2\pi TD_e}{v_B^2}\geq 1=:\mathcal{B}_e
\end{equation}
which is in agreement with the findings in~\cite{Blake:2016sud}. }

\subsection{Shear viscosity to entropy density ratio via a (weaker) horizon formula}

To compute the \emph{shear viscosity-to-entropy density} ratio, ${\eta}/\mathcal{S}$, we employ the method devised in~\cite{Hartnoll:2016tri}. {We focus on the bulk metric perturbation $g_{xy}=\epsilon r^2 h(r)e^{-i\omega t}$ of the eigenmode type about the black hole background \eqref{eq:f4dyon}. However, the full set of linearized field equations can be found in appendix~\ref{sec:LFE}. We are interested in the $h_x^y\equiv h$ mode which, having set $k=0$, decouples from the other fluctuations. It is determined by the second order ODE, $\mathcal{G}_{xy}=0$, which can be written as}
\begin{equation}
\frac{\partial_r(r^2F\partial_r h)}{r^2}+\pqty{\frac{\omega^2}{F}-m(r)^2}h=0.\label{eq:waveeq}
\end{equation}
The explicit expression of the (effective) mass function will be stated at a later point. The shear viscosity is then computed in terms of the correlator:
\begin{equation}
{\eta}=\lim\limits_{\omega\to 0}\frac{1}{\omega}\Im G^{R}_{T^{xy}T^{xy}}(\omega,k=0)= r_0^2h_o(r_0)^2={4 G_{\mathrm{eff}}\mathcal{S}h_o(r_0)^2},
\end{equation}
where $h_o$ is the solution to \eqref{eq:waveeq} at zero frequency, $\omega=0$, which (i) is regular at $r_0$ and (ii) goes like unity near radial infinity. Then, the \emph{shear viscosity-to-entropy density} ratio is 
\begin{equation}
\frac{{\eta}}{\mathcal{S}}={4 G_{\mathrm{eff}}h_o(r_0)^2}.
\end{equation}
Obviously, when $\alpha=0$, the entropy reduces to $\mathcal{S}_o$, and $G_{\mathrm{eff}}=(16\pi)^{-1}$; the expression for the ratio assumes the standard form~\cite{Hartnoll:2016tri}. {Moreover, since the metric fluctuation is massive, and assuming a positive effective mass squared, we know that $h_o(r_0)<1$ which follows from a simple argument, deliberately illustrated in~\cite{Hartnoll:2016tri}.} Since $G_{\mathrm{eff}}<1$ strictly for nontrivial $\alpha$, we already know that the simple $(4\pi)^{-1}$ bound is definitely violated at finite temperatures. The mass squared is given by 
\begin{equation}
m^2=\frac{p}{r^2}\pqty{\hat{\eta}p+\frac{2\sqrt{2}\alpha r(3r-F')}{2r+\sqrt{2}\alpha p}},\label{eq:efmas}
\end{equation}
where again $F$ and its derivatives are understood to be on the background shell. {First of all, we observe that as $\alpha\to 0$,  we recover the standard mass squared term, $\hat{\eta} (p/r)^2$, resulting from an EMA bulk.} Then, \eqref{eq:efmas} will be strictly positive at $r_0$. It will be also finite positive in the $T\to 0$ limit where the black hole becomes extremal with $r_0=r_*$. There is no general argument why $m^2$ needs to be positive in general, but in our case it so happens that it is a strictly positive function of the radial coordinate in the physical domain of interest $r_0<r<\infty$. 

Now, we define $b=p/r_0$, and we notice that~\eqref{eq:waveeq} with $\omega=0$ has already terms linear in $b$; indeed, it reads 
\begin{equation}
\sum_{n=0}^{3}(\sqrt{2}\alpha bz)^nf_n(z)=0,\label{eq:ODE}
\end{equation}
where
\begin{eqnarray}
f_0(z)&=&z^2[2-b^2z^2+(b^2-2)z^3]h''(z)+z[(b^2-2)z^3-4]h'(z)-2b^2z^2 h(z),\\
f_1(z)&=&z^2[2-b^2z^2+(b^2-2)z^3]h''(z)+\frac{z}{2}[b^2z^2+(b^2-2)z^3-10]h'(z)-\nonumber\\
&&-[2+3b^2z^2+(b^2-2)z^3]h(z),\\
f_2(z)&=&\frac{z^2}{4}[2-b^2z^2+(b^2-2)z^3]h''(z)+\frac{z}{4}(b^2z^2-6)h'(z)-(2b^2z^2+3)h(z),\\
f_3(z)&=&-\frac{b^2z^2+6}{4}h(z).
\end{eqnarray}
Here, a convenient change of the radial coordinate, $z=r_0/r$, was performed, such that the horizon and boundary are located at $z=1$ and $z=0$, respectively. Also, $\hat{\eta}$ was set to unity for convenience. We observe that, due to the new coupling, there are odd powers of $b$ introduced in the differential equation. Treating $b$ perturbatively, if we were to expand the solution as 
\begin{equation}
h_o(z)=\sum_{n=0}^{\infty}b^{2n}h_{o,2n}(z),
\end{equation}
plugging it back into \eqref{eq:ODE}, and solving the ODE order by order, we would find inconsistencies already at order $b$. Ergo, a general expansion of the form
\begin{equation}
h_o=\sum_{n=0}^{\infty}b^{n}h_{o,n},\label{eq:series}
\end{equation}
is necessary. All we need to do now, is solve order by order. We also remind the reader that we work at zero chemical potential. At zeroth order we need to obtain a solution to
\begin{equation}
z(z^3-1)h_{o,0}''+(z^3+2)h'_{o,0}=0
\end{equation}
where a prime denotes differentiation with respect to $z$. The general solution to this reads 
\begin{equation}
h_{o0}=c_2-\frac{c_1\ln(1-z^{3})}{3}.
\end{equation}
Regularity at the horizon suggests that $c_1$=0, while the boundary conditions at $z=0$ imply that $c_2=1$. Hence, $h_{o,0}=1$, and we need to make all other $h_{o,n}$, $n>0$, vanish at $z=0$ so that the asymptotic behavior of $h_o$ meets condition (ii). Moving on to linear order in $b$, we find that the solution to
\begin{equation}
z^2(z^3-1)h''_{o,1}+z(z^3+2)h'_{o,1}-\sqrt{2}\alpha z(z^3-1)=0,
\end{equation}
compatible with the aforementioned conditions, reads
\begin{equation}
h_{o,1}=\frac{\alpha}{6\sqrt{2}}\bqty{\sqrt{3}\pi+12 z- 6\sqrt{3} \mathrm{atan}\pqty{\frac{1+2z}{\sqrt{3}}}-9\ln(1+z+z^2)}
\end{equation}
At second order, the solution is already too lengthy to write down. It involves many dilogarithmic and arctangent functions. All contributions due to the nonminimal coupling go as $\alpha^2$ and schematically, 
\begin{equation}
h_{o,2}(1)\sim \frac{1}{18}\pqty{\sqrt{3}\pi - 9 \ln 3}+\alpha^2(. . .).
\end{equation}
Higher orders, $k$, will go as the solutions in~\cite{Hartnoll:2016tri} plus $\alpha^{k}$ corrections if $k$ even, while if $k$ is odd, the solution will have an overall $\alpha^{k}$ factor, such that when we switch off the coupling constant we recover the $h_o$ of the linear axion model. 

\begin{figure}[ht!]
\centering
\subfigure[]{
\includegraphics[width=.45\textwidth]{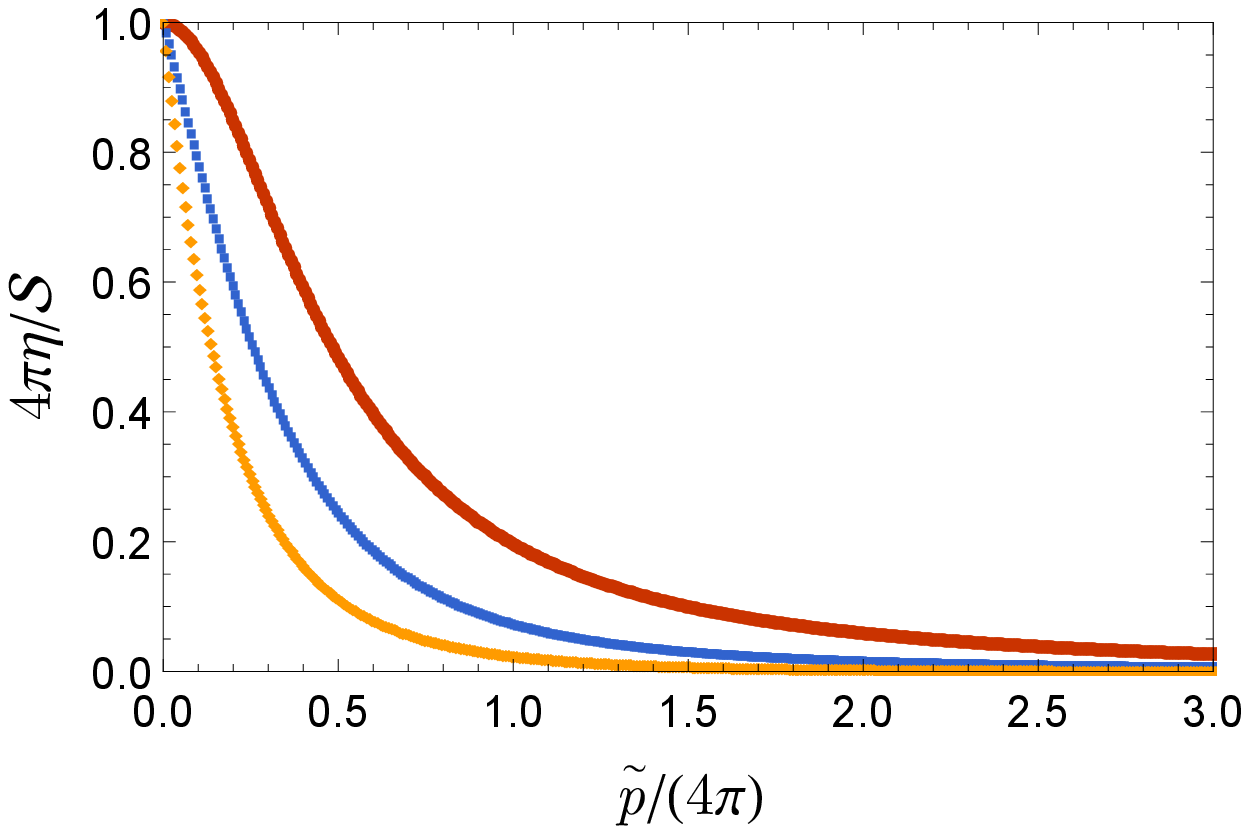}
}
\subfigure[]{
\includegraphics[width=.45\textwidth]{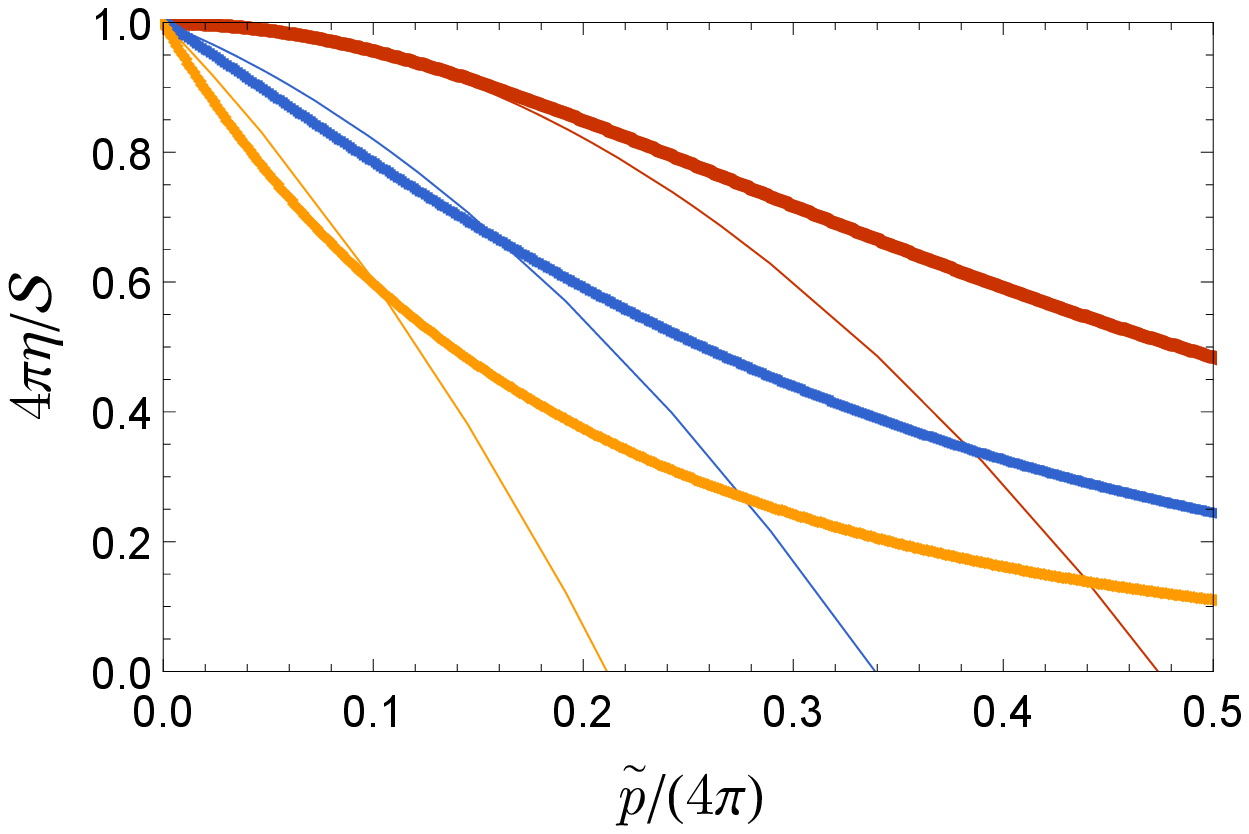}
}

\caption{In subfigure (a) we exhibit $4\pi\eta/\mathcal{S}$ vs. $\tilde{p}/(4\pi)$ by using numerical methods to solve~\eqref{eq:ODE}. The thick red line stands for the ratio in the linear axion model, the thick blue line is for $\alpha=0.2$, and the thick yellow one is for $\alpha=0.5$. In subfigure (b) we compare the numerical results to the high temperature approximations, up to order $\tilde{p}^2$. Thick lines are as in subfigure (a), whereas the thin ones depict the approximation with the color pattern being the same as in subfigure (a).}
\label{fig:etasnum}
\end{figure}

Just as a minor example, let us try a very crude and inelegant approximation of the ${\eta}/\mathcal{S}$ ratio in the high temperature regime. We find that 
\begin{eqnarray}
\frac{{4\pi\eta}}{\mathcal{S}}&=&1-\frac{\alpha(\sqrt{3}\pi+9\ln 3-6)}{4\sqrt{2}\pi}\tilde{p}+\bqty{\frac{\sqrt{3}\pi\!-\!9\ln 3}{16\pi^2}\!+\!\frac{3\alpha^2(. . .)}{128\pi^2}}\tilde{p}^2+\mathcal{O}\pqty{\tilde{p}^3}. \label{eq:etas}
\end{eqnarray}
Unfortunately, this approximation, being so crude, is not very helpful; a numerical solution to~\eqref{eq:ODE} is certainly necessary in order to obtain a better insight. This can be seen in figure~\ref{fig:etasnum}. In subfigure (b), for the approximation plots corresponding to the cases with nonvanishing $\alpha$, we have neglected the $\alpha^2$ contributions in~\eqref{eq:etas}, in order to avoid using information that is not explicitly displayed here. Since this contribution comes with an overall negative sign in the end, it is, of course, expected that the curves fit even better to the numerical results in the very small $\tilde{p}$ region where the approximation applies. From subfigure (a), we observe that, apart from a manifestly more brutal violation of the KSS bound, the ratio exhibits a qualitatively similar behavior at both  extremes, always in comparison to the results obtained in the case of the linear axion model. It goes to unity as $\tilde{p}\to 0$, while it tends to zero when $\tilde{p}\to \infty$. In general, the violation itself is nothing really surprising,\footnote{See also~\cite{Baggioli:2020ljz} for a very recent discussion on bounds when translation symmetry is broken.} mainly because the shear mode mass is everywhere nonvanishing. Albeit that, it is nevertheless interesting to visualize the impact of the particular Horndeski deformation on the ratio under study, which, in the case of broken translation symmetry, does not have the usual hydrodynamic interpretation, but rather corresponds to the entropy production rate. 

\section{Concluding remarks\label{sec:conc}}

In this work, we have started by considering a specific model of Horndeski gravity with $G_3=0=G_5$. Inspired by~\cite{Babichev:2017guv}, we took $(D-2)$-many copies of this model such that the massless scalar fields, $\psi^{I}=p\delta^{I}_ix^{i}$, are homogenesouly distributed along the $(D-2)$-many planar directions; the final action can be intuitively expressed as Einstein gravity with a running effective gravitational coupling, plus matter fields with higher derivatives, accompanied by a Maxwell term. By doing so, we managed to construct novel charged planar black holes with nontrivial axionic hair. We studied the horizon structure of the four-dimensional solution with AdS asymptotics, revealing a mass region in which the black holes possess two horizons which coalesce into one in the extremal case. Note that since the extremal mass can be negative, it is possible to have black holes with negative mass, as well. {We argued that there exist solutions in a certain parameter window --- where the kinetic terms acquire the ``wrong'' sign ---, which do not lead to a violation of the weak energy condition. However, this option was neglected on purpose, by means of a limiting argument}. A near-horizon $\mathrm{AdS}_2\times \mathds R^2$ structure was observed, whereas these solutions tend to standard unit-radius $\mathrm{AdS}_4$ at asymptotic infinity. A straightforward dyonic extension of these black holes was given, while we also exhibited the three-dimensional solution which does not flow from limiting the $D$-dimensional result, and requires separate integration of the field equations. {In order to study some holographic aspects of the model, we proceeded by investigating the thermodynamic properties of~\eqref{eq:f4} where the entropy was derived via two routes: first, we used Wald's Noether charge entropy formula, and second, we employed the conventional Euclidean path integral approach, with the reference spacetime being the extremal solution at zero electric charge}. Contrary to the discrepancies advertised in the beginning of section~\ref{sec:thermo}, associated with the toy model in~\cite{Anabalon:2013oea}, we found that, in our case, both methods agree on the result. {As expected for a nonminimally coupled scalar-tensor theory, the 1/4-area law for the entropy is modified, although it can be said to hold in units where~\eqref{eq:kappaef} equals unity. In this sense, and to some extent, hints to the 1/4-area law are still there. Expressions for the mass and the Hawking temperature were also provided, and the first law was shown to hold true, provided that $p$ is held fixed. }

Next, we used the powerful method devised in~\cite{Donos:2014cya,Donos:2013eha} to compute the linear thermoelectric DC response of the holographic dual system to some external electric field and some thermal gradient. This was done by means of black hole horizon data only, exploiting the radial conservation of the electric/heat currents. Analytic expressions were found for these currents, along with a detailed derivation of the DC transport matrix. As a consistency check, we verified that the matrix was symmetric --- a consequence of invariance under time reversal ---, while we found that the $\bar{\kappa}/\upalpha$ ratio stated in~\cite{Donos:2014cya} was modified, becoming~\eqref{eq:ka}. The comparison is always carried out with respect to the linear axion model~\cite{Bardoux:2012aw,Andrade:2013gsa}. We saw that the behavior of the electric conductivity at the two termperature extremes was not altered by the presence of the Horndeski coupling, a fact suggesting that it is rather governed by the choice of electrodynamics instead; {it would be interesting to consider different types of non-linear electrodynamics (see for example refs.~\cite{Born:1934gh,Cisterna:2020rkc,Liu:2019rib,Hassaine:2007py,Cremonini:2017qwq}) coupled to this gravitational toy model, or, even better, axion-gauge sector couplings, investigating how the parameters mingle with each other, and if it is possible to flow towards an insulating phase at strong dissipation.}

Knowing the linear responses of the system, we computed the thermal conductivity $\kappa$ at zero electric current. {Quite remarkably, we found a parametric violation of the $4\pi^2/3$ lower bound for the ratio $\kappa/T$; instead of a universal number, we were led to a bounding function with a global minimum at $p\sqrt{\hat\eta}=\mu$. At fixed chemical potential, the bounding function is driven by the dissipation scale and the strength of the gravity bulk deformation. As $\alpha/\tilde{\mu}$ grows larger, the function itself becomes steeper, with the lowering being most effective at dissipation scales close to the order of $\tilde{\mu}$. To the best of our knowlsedge, we are not aware of other works reporting a similar result. It would be interesting to scrutinize the causal relation between the proposed deformation and the bound violation, in an effort to unveil a possible connection between bulk modifications and the bounding function.} Moving on, the upper bound,  $\kappa/T\leq \mathcal{S}^2/Q_e^2$, was still found to hold good. The Kelvin formula was also verified, in favor of the argument~\cite{Blake:2016jnn,Davison:2016ngz} that its validity is associated with the flow of holographic models towards an $\text{AdS}_2\times \mathds{R}^2$ fixed point in the IR.

Next, we considered a generalized version of Einstein's relation where the diffusitivities are mixed by a term $\mathfrak{M}$~\cite{Hartnoll:2014lpa}, with the constants describing the coupled diffusion being $D_{\pm}$. {We explicitly showed that, for this model, the mixed diffusion constants do indeed decouple in the incoherent regime where the mixing is suppressed. In this regime, the charge and energy diffusitivities can be used, instead . As also observed in~\cite{Kim:2017dgz}, the mixing becomes maximal for small $\tilde{\mu}$, in the sense that there is an opposite identification of $D_{\pm}$ with $D_c$ and $D_e$, opposite to the way these quantities are matched in the incoherent phase. The deformation did not affect the decoupling process, neither was the new coupling present in the leading orders of the incoherent expansion of the diffusion constants.} Considering the Blake refinement~\cite{Blake:2016wvh} of the $T D/v^2$ lower bound --- originally conjectured in~\cite{Hartnoll:2014lpa} ---, where $v=v_B$ is the butterfly velocity, we calculated the latter only to find out that the proposed modification did not alter the so-called shift equation drastically, nor did it have any impact on the screening length $m$. The velocity obeyed the generic formula obtained for a pure Einstein bulk with minimally-coupled matter~\cite{Blake:2016wvh}; the new coupling entered the velocity only through the explicit expression of the horizon, eq.~\eqref{eq:r0temp}. Since the incoherent expansion of $r_0$ is independent of $\alpha$ at leading order, we concluded and graphically demonstrated that $T D_{c(e)}/v_B^2$ is eventually bounded by the standard numbers from below.

Next, we employed the weaker horizon formula~\cite{Hartnoll:2016tri} to determine the \emph{shear viscosity-to-entropy density} ratio, $\eta/\mathcal{S}$, at zero chemical potential. Since the $g_{xy}$ fluctuations were massive, with a positive effective mass squared given by~\eqref{eq:efmas}, it was no surprise that the model led to a violation of the simple $(4\pi)^{-1}$ bound. Performing a very crude approximation, we noticed that, at small $\tilde{p}$, the Horndeski deformation did not only contribute to the even-power subleading terms of the ratio expansion, but allowed for odd-power corrective terms as well, the latter entirely imputed to the presence of the nonminimal coupling. We also managed to solve~\eqref{eq:ODE} numerically, displaying $\eta/\mathcal{S}$ in figure~\ref{fig:etasnum}. In comparison to the linear axion model, a more brutal violation of the KSS bound was observed. {However, the behavior at both dissipation extremes was found to be similar with the ratio going to zero (unity) when $\tilde{p}\to \infty$ ($\tilde{p}\to 0$). We note here that since translation symmetry is broken, the $\eta/\mathcal{S}$ ratio cannot have the standard hydrodynamic interpretation; it is rather associated with the entropy production rate. To that extent, as stated in~\cite{Baggioli:2020ljz}, it is plausible that another ratio makes more sense when boundary momentum gets relaxed. The idea is to look for a lower bound of the momentum diffusion constant, the claim being that this bound may match the KSS one, although $4\pi\eta/\mathcal{S}\leq 1$. However, calculating the momentum diffusion constant, requires the knowledge of the correlator $G^R_{T^{ty}T^{ty}}$ with $\omega,k\neq 0$ (if momentum was set along the $x$ direction), which is determined by the shear metric fluctuation $h^y_t$. As pointed out in appendix~\ref{sec:LFE}, one must decouple the $ty$ mode from the rest, in order to obtain a clean ODE for it, which can then be solved perturbatively. Unfortunately, although the decoupling is carried out as a relatively straightforward task in the case of the linear axion model~\cite{Ciobanu:2017fef}, here, we were not able to find a way to separate this mode. Therefore, we believe that a tiresome numerical approach is favored, which, however, lies out of the scope of this paper.} Finally, we mention here, as a possible further development, that it is appealing to consider \eqref{eq:f4dyon} as the background solution, instead, and work out the bounds at finite magnetic field, whereas the magneto-transport properties of the model is another topic, interesting in its own right. As a closing remark, we note that it would be very interesting to further investigate this model in the relevant context of~\cite{Li:2018kqp,Li:2018rgn}.

\acknowledgments
The authors would like to thank Adolfo Cisterna, Mokhtar Hassaine and Julio Oliva for their time and contributions during the early stages of this work. They would also like to thank Anastasios Petkou for valuable comments and suggestions, while they are also grateful to the referees, for highlighting important points and making beneficial suggestions. K.P acknowledges financial support provided by the European Regional Development Fund through the Center of Excellence TK133 ``The Dark Side of the Universe'' and PRG356 ``Gauge gravity: unification, extensions and phenomenology''. J.F acknowledges the financial support from ANID through the Fellowship 22191705. 

\appendix
\section{Linearized field equations}\label{sec:LFE}
Let us consider the solution~\eqref{eq:f4}. A change of variables $r\to 1/ \mathcal{R}$ with $dr^2=\tfrac{1}{\mathcal{R}^4}d{\mathcal{R}}^2$, allows us to write the metric as 
\begin{equation}
    ds^2=\frac{1}{\mathcal{R}^2}\pqty{-G(\mathcal{R})dt^2+\frac{1}{G(\mathcal{R})}d\mathcal{R}^2 +dx^2 + dy^2},
\end{equation}
where 
\begin{equation}
    G(\mathcal{R}):=\frac{1}{\mathfrak f(\mathcal{R})}\bqty{1-\frac{\hat\eta p^2\mathcal{R}^2}{2}-\pqty{\frac{\mathcal{R}}{\mathcal{R}_0}}^3\pqty{1-\frac{\hat\eta p^2\mathcal{R}_0^2}{2}}},
\end{equation}
for $Q_e=0$ with
\begin{equation}
    \mathfrak{ f}(\mathcal{R}):=1+\frac{\alpha p \mathcal{R}}{\sqrt{2}}.
\end{equation}
In what follows it will be convenient to rescale $\mathcal{R}$ as $z=\mathcal{R}/\mathcal{R}_0$ such that $z= 0$ corresponds to the conformal boundary whereas $z=1$ indicates the location of the horizon. Now, we turn on a small source on the boundary which amounts to perturbing our background solutions as 
\begin{equation}
    \delta g_{ty}= h^y_t g_{yy}e^{-i(\omega t-kx)},\quad \delta g_{xy}= h^y_x g_{yy}e^{-i(\omega t-kx)},\quad \delta \psi^2= \Psi e^{-i(\omega t-kx)},
\end{equation}
where $h,\Psi$ are functions of $z$. {At the level of the linearized field equations, the shear metric modes will in general be coupled to the scalar fluctuation $\delta \psi^2$. Moreover, we have now set the momentum along the $x$ direction, but we could have equally set it along the $y$ direction}, as we did in the bulk of this paper. The linearized equations of motion read
\begin{align}
    \frac{z^2}{\tilde{r}_0^2}\partial_z\pqty{\frac{\mathfrak f {h^y_t}'}{z^2}}-\pqty{i\omega\Psi+p h^y_t}\frac{{\mathfrak g}}{G}-\frac{k}{G}\pqty{\omega h^y_x+kh^y_t}&=0,\label{eq:h14}\\
    \frac{z^2}{\tilde{r}_0^2}\partial_z\pqty{\frac{G {h^y_x}'}{z^2}}\mathfrak f+\frac{\omega}{G}\pqty{kh^y_t+\omega h^y_x}\mathfrak{f}+\pqty{ik\Psi-ph^y_x}\mathfrak{h}&=0,\label{eq:h34}\\
    i\omega \mathfrak f {h^y_t}'+ik G {h^y_x}'-\mathfrak g G\Psi'&=0,\label{eq:h24}\\
    \frac{z^2}{\tilde{r}_0^2}\partial_z\pqty{\frac{\mathfrak g G\Psi'}{z^2}}\mathfrak f+\frac{\Psi}{G}\pqty{\omega^2\mathfrak g\mathfrak f-k^2G \mathfrak h}-\frac{ip}{G}\pqty{\omega h^y_t \mathfrak g\mathfrak f+kh^y_xG\mathfrak h}&=0,
\end{align}
where $\mathfrak{f,g,h}$ are now meant as functions of $z$ with
\begin{align}
    \mathfrak g(z)&:=\hat\eta p+\alpha\frac{{2G-zG'}}{\sqrt{2}\mathcal{R}_0 z},\\
    \mathfrak h(z)&:=\hat\eta p+\alpha \frac{6+\hat\eta p^2\mathcal{R}_0^2z^2-2\pqty{2G-zG'}}{\sqrt{2}\mathcal{R}_0 z}.
\end{align}
Observe that by dividing \eqref{eq:h34} by $\mathfrak f$ and setting $k=0$, the $h^y_x$ mode, determining the correlator $G^R_{T^{xy}T^{xy}}$, is fully decoupled. Then, reverting back to the original radial coordinate $r$, one obtains \eqref{eq:waveeq}. On the other hand, the correlator $G^R_{T^{ty}T^{ty}}$ is determined by the $h^y_t$ mode which does not fully decouple from the remaining fluctuations; in the case of the linear axion model~\cite{Ciobanu:2017fef}, that is $\alpha=0$, one would take the derivative of \eqref{eq:h14} with respect to $z$, followed by the use of \eqref{eq:h24}, in order to finally write \eqref{eq:h14} as a third-order ODE for $h^y_t$. Here, because of the new coupling, such a strategy won't work when $\omega,k\neq 0$.

\bibliographystyle{JHEP}
\bibliography{Refs}

\end{document}